\documentclass[onecolumn]{emulateapj}

\slugcomment{Accepted to ApJ, in press}

\shortauthors{Rogers 2015}

\usepackage{amsmath}
\usepackage{graphicx}
\usepackage{epstopdf}
\usepackage{epsfig}
\usepackage{natbib}
\usepackage{wasysym}
\usepackage{lipsum}
\usepackage{mathrsfs}

\usepackage{color}

\begin{document}

\title{Most 1.6 Earth-Radius Planets are not Rocky}
\author{Leslie A. Rogers\altaffilmark{1, 2, 3} }

\altaffiltext{1}{Department of Astronomy, California Institute of Technology, Pasadena, CA 91125, USA}
\altaffiltext{2}{Department of Geophysics and Planetary Sciences, California Institute of Technology, Pasadena, CA 91125, USA}
\altaffiltext{3}{Hubble Fellow}

\begin{abstract}
The {\it Kepler} Mission, combined with ground based radial velocity (RV) follow-up and dynamical analyses of transit timing variations, has revolutionized the observational constraints on sub-Neptune-size planet compositions. The results of an extensive {\it Kepler} follow-up program including multiple Doppler measurements for 22 planet-hosting stars more than doubles the population of sub-Neptune-sized transiting planets that have RV mass constraints. This unprecedentedly large and homogeneous sample of planets with both mass and radius constraints opens the possibility of a statistical study of the underlying population of planet compositions.  We focus on the intriguing transition between rocky exoplanets (comprised of iron and silicates) and planets with voluminous  layers of volatiles (H/He and astrophysical ices). Applying a hierarchical Bayesian statistical approach to the sample of {\it Kepler} transiting sub-Neptune planets with Keck RV follow-up, we constrain the fraction of close-in planets (with orbital periods less than $\sim50$~days) that are sufficiently dense to be rocky, as a function of planet radius. We show that the majority of $1.6~R_{\oplus}$ planets are too low density to be comprised of Fe and silicates alone. At larger radii, the constraints on the fraction of rocky planets are even more stringent. These insights into the size demographics of rocky and volatile-rich planets offer empirical constraints to planet formation theories, and guide the range of planet radii to be considered in studies of the occurrence rate of ``Earth-like" planets, $\eta_{\oplus}$. 
\end{abstract}

\section{Introduction}

After subsisting on the major bodies orbiting the Sun for centuries, astronomers are now placing the Solar System into context with the discovery of a plethora of exoplanets. 
Exoplanets detected both in transit and by their dynamical influence are very valuable. The radius derived from the transit depth and the mass derived from radial velocity (RV) measurements or transit timing variations (TTVs), together give the planet density and some handle on the planet composition. To date, more than 200 transiting planets have measured masses \citep[e.g., exoplanets.org,][]{WrightEt2011PASP}. 
The accumulating census of transiting planets with measured masses contains information about the underlying composition distribution of planets and about the masses of the more than 3000 {\it Kepler} transiting planet candidates \citep{BoruckiEt2011ApJ, BatalhaEt2013ApJ} that currently lack dynamical confirmation.

Planet mass-radius measurements are especially important for planets that are smaller than Neptune $\left(R_p\lesssim 4 R_{\oplus}\right)$. For planets in this size range, a wide diversity of planet compositions are a priori plausible; rock, astrophysical ices (H$_2$O, NH$_3$, and CO), and H/He gas can all make significant contributions to both the planets' mass and volume \citep[e.g.,][]{Rogers&Seager2010ApJ, Rogers&Seager2010bApJ}. 
In the Solar System, there are no planets with masses and radii intermediate between the Earth and the ice giants. As a statistical sample of sub-Neptune-sized exoplanets with measured masses and radii accumulates, we may begin to study the intriguing transition between planets that are predominantly rocky and those with voluminous layers of volatiles (astrophysical ices and H/He).  

The {\it Kepler} Science Team targeted 22 {\it Kepler} Objects of Interest (KOIs) hosting planet candidates with $R_p<4~R_{\oplus}$ in an extensive Keck HIRES RV follow-up program \citep{MarcyEt2014ApJS}. 
 The planet candidates were selected for RV follow-up based on their quiet host stars ({\it Kepler} Magnitude $K_p<13.5$, $T_{\rm eff}<6100~\rm{K}$, $v\sin{i}<5~\rm{km\,s^{-1}}$) and detectable predicted RV-amplitudes $\left(K>1~\rm{m\,s^{-1}}\right)$. Each target received 20--50 RV measurements between 2009 July and 2013 August. A mean velocity precision of $\sim2~{\rm m\, s^{-1}}$ was achieved for the planet hosts, which have {\it Kepler} magnitudes ranging from 8.77 to 13.5. Orbital fits and MCMC analyses were employed to derive planet masses and mass upper limits from the Doppler RVs, yielding mass constraints for 49 planets (42 transiting and 7 non-transiting) in 22 planetary systems. Sixteen transiting planets have mass measurements at a confidence level of $2\sigma$ or better, while the rest have marginal RV detections or mass upper limits. The \citet{MarcyEt2014ApJS} survey more than doubles the number of known sub-Neptune-size transiting planets with RV mass constraints.  

The {\it Kepler} transiting planets with Keck HIRES RV follow-up is the largest and most homogeneous sample of sub-Neptune-size planet mass-radius measurements, to date. It opens an unprecedented window into the demographics of small planet bulk compositions.  \citet{Weiss&Marcy2014ApJL} fit these measured masses and radii (along with those of an additional 9 planets smaller than $4~R_{\oplus}$ with masses vetted on exoplanets.org) to power-law relations and found a nearly linear mass-radius relation: $M_p/M_{\oplus}=2.69\left(R_p/R_{\oplus}\right)^{0.93}$.

In this work, we focus on the intriguing threshold between rocky planets and exo-Neptunes with voluminous gas layers, a transition which has implications for planet formation, evolution, habitability, and the interpretation of the {\it Kepler} planet radius distribution. Instead of fitting simple functional forms to the measured properties of planets, we apply statistical tools to investigate how the HIRES-RV and {\it Kepler} transit depth measurements constrain the underlying composition distribution of planets. We apply a hierarchical Bayesian analysis to constrain, as a function of planet size, the fraction of planets that are sufficiently dense to be rocky. As inputs, our approach takes samples from the mass-radius posterior distributions output from fitting the RV and transit photometry data. Our approach naturally accounts for both the non-gaussian likelihoods, and the significant correlations between the RV-measured masses of planets in multi-planet systems \citep[neither of which are taken into account by fitting to estimate outputs or summary statistics, as in][]{Weiss&Marcy2014ApJL}. 

We present updated mass-radius measurements for small planets and describe our sample of planets in Section~\ref{sec:MR}. We outline our statistical approach to constraining the fraction of planets that are sufficiently dense to be rocky, as a function of planet size in Section~\ref{sec:stat}. We present the results of our analysis in Section~\ref{sec:res}, and then discuss and conclude in Sections~\ref{sec:dis} and \ref{sec:con}.

\section{Properties of Individual Sub-Neptune-Size Planets}
\label{sec:MR}

\subsection{Defining a Statistical Sample of Planets}
\label{sec:statMR}

The full collection of planets having constraints on both their mass, $M_p$, and radius $R_p$ (Figure~\ref{fig:MRall}) is a heterogeneous sample. Some planets were initially discovered from ground-based RV surveys, and later found to transit (GJ~436b, 55 Cnc e, HD~97658b). A larger sample of planets were initially discovered by transit surveys, and then confirmed with RV follow-up (e.g., GJ~1214b, CoRoT-7b, HAT-P-11b). The {\it Kepler} transit survey alone has made a tremendous contribution to populating the mass-radius diagram of small planets. The {\it Kepler} discoveries themselves are a diverse sample; many have mass constraints derived from TTVs (e.g., Kepler-11b,c,d,e,f,g, Kepler-30b,c, Kepler-36b,c, and the \citet{Wu&Lithwick2013ApJ} planets), while the rest have RV-derived mass constraints (e.g., Kepler-4b, Kepler-10b,c, Kepler-19b, Kepler-20b,c,d, Kepler-21b, Kepler-22b, KOI-94b, and the \citet{MarcyEt2014ApJS} planets) or joint RV-TTV constraints (e.g., Kepler-18b,c,d).  

For our statistical study of the planet composition distribution, we focus on the {\it Kepler} planets and planet candidates with RV-constrained masses. Each planet transit and RV survey has its own set of selection effects and biases. 
 By restricting our sample to the {\it Kepler} planets with Keck HIRES RV follow-up, we work with the largest, most homogeneous sample of sub-Neptune planet mass-radius measurements collated to date. In addition to the planet masses published in \citet{MarcyEt2014ApJS}, we also include planetary systems with previously published Keck HIRES RVs: KOI-70 (Kepler-20), KOI-72 (Kepler-10), KOI-84 (Kepler-19),  KOI-87 (Kepler-22), and KOI-975 (Kepler-21). 
  We have chosen not to include {\it Kepler} planets with TTV-constrained masses in our sample, since they are subject to separate selection effects \citep[which may preferentially favor low-density planets, e.g.,][]{JontofHutterEt2014ApJ, Weiss&Marcy2014ApJL}.  
 
The sample of sub-Neptune-size {\it Kepler} planets with Keck HIRES RV follow-up includes planets on close-in orbits. All but 4 of the planets in our sample have orbital periods less than 50~days, while all but 1 of the planets with RV-measured masses more than $2~\sigma$ above zero have orbital periods below 16.2~days. Kepler-22b (KOI-87.01) is the long period outlier in the sample, with an orbital period of 290~days, and a marginal $\sim2~\sigma$ mass measurement of $32^{+10}_{-14}~M_{\oplus}$ (including RV measurements up to 2013 August). The planets in our sample receive bolometric incident flux, $F_p$, between 1.1 and 3700 times that received by the Earth at 1AU from the Sun, $F_{\oplus}$. 

For each {\it Kepler} planet candidate selected for RV follow-up, the measured planet mass provides an unbiased sampling of close-in planet masses for its particular planet radius. The selection of KOIs for Keck-HIRES follow-up \citep[as detailed in ][]{MarcyEt2014ApJS} was neither random, nor fully algorithmic. As the {\it Kepler} team's priorities shifted toward characterizing smaller and smaller planets as the mission progressed, the selection criteria prioritizing KOIs for follow-up observations also evolved. The planets were selected for Keck HIRES RV follow-up based on their planet radius and stellar properties, and then pursued with Keck HIRES in a mass-blind way. We factor out dominant selection effects by focussing on the conditional distribution of masses for close-in planets as a function of planet radius. We plan to incorporate a more elaborate treatment of selection effects in our hierarchical Bayesian model in future work.  

\subsection{Measuring $R_p$ and $M_p$ from Transit and RV Data}
\label{sec:MRfit}

For each planet host in our sample, the {\it Kepler} photometry and Keck RVs were simultaneously fit with an analytic model for transiting planets on Keplerian non-interacting circular orbits. The stellar properties are constrained either by analysis of high-resolution Keck reconnaissance spectra with SME \citep[Spectroscopy Made Easy][]{Valenti&Piskunov1996A&AS}, or by asteroseismology analysis. There was no strong evidence in the RVs for eccentric planet orbits, though 7 of the planet host stars did have RV trends indicative of non-transiting planets. The fitting procedure is described in detail in \citet{MarcyEt2014ApJS}. This full photo-dynamical analysis was performed in 2013 May. To incorporate RV measurements from the 2013 observing season, the analysis of the Keck RVs was repeated in Fall 2013. This time, however, only the RV amplitude of each planet was fit, with all other system parameters held fixed \citep{Isaacson2013PC}. 

In our statistical analysis we use the planet mass distributions obtained from this updated fit to the RVs. We note that some of the masses used in this work maybe slightly different from those quoted in Table~2 of \citet{MarcyEt2014ApJS}, which mixes the two different procedures to fit the light curve and RV data (from Spring 2013 and Fall 2013). For our statistical study, we apply the same fitting procedure consistently to all 22 planet hosts in the \citet{MarcyEt2014ApJS} and the 5 previously published small-planet-hosting stars with Keck-HIRES RVS. The difference in the masses obtained by the two procedures is no more than $1.4~\sigma$ for any planet. 

The MCMC chains sampling the masses of the planets orbiting each star are directly incorporated into our statistical analysis of the planet composition distribution. In this way, we fully account for correlations between the RV masses of planets orbiting the same star, which may be induced both by the mutual dependence of the multi-planets' properties on the stellar properties and by correlations in the planets semi-amplitudes obtained from fitting the RVs. Accounting for correlations is especially important in the case of marginal RV detections, for which the posterior distribution of the planet mass is most strongly subject to correlations with the masses of other planets in the system. Pairs of planets with strong correlations in their measured masses (correlation coefficients $\left|R\right|>0.1$) include KOIs 70.01 and 70.02 $\left(R=-0.31\right)$, 70.01 and 70.03 $\left(R=-0.13\right)$, 70.02 and 70.03 $\left(R=0.15\right)$, 116.01 and 116.02 $\left(R=0.11\right)$, 148.01 and 148.02 $\left(R=-0.35\right)$, 148.01 and 148.03 $\left(R=0.14\right)$, 148.02 and 148.03 $\left(R=-0.18\right)$, 245.02 and 245.03 $\left(R=-0.16\right)$, and 246.01 and 246.02 $\left(R=-0.12\right)$.

We have neglected correlations among the planet radii and any correlations between a planet's radius and its mass, since the RV data were fit independently of the transit light curve data. We approximate the distribution of planet radii as gaussian with the mean and standard deviation given in Table~2 of \citet{MarcyEt2014ApJS}. Since the radii of the planets in our sample are far more tightly constrained than the masses, correlations affecting the planet radii are expected to be minor and to have a subdominant effect on our results. This has borne out in the results of the full photo-dynamical fit to the {\it Kepler} photometry and a subset of the RV data \citep{MarcyEt2014ApJS}. 

\subsection{Identifying Potentially Rocky Planets}
\label{sec:rockyThresh}

\begin{figure}
\centering
\includegraphics{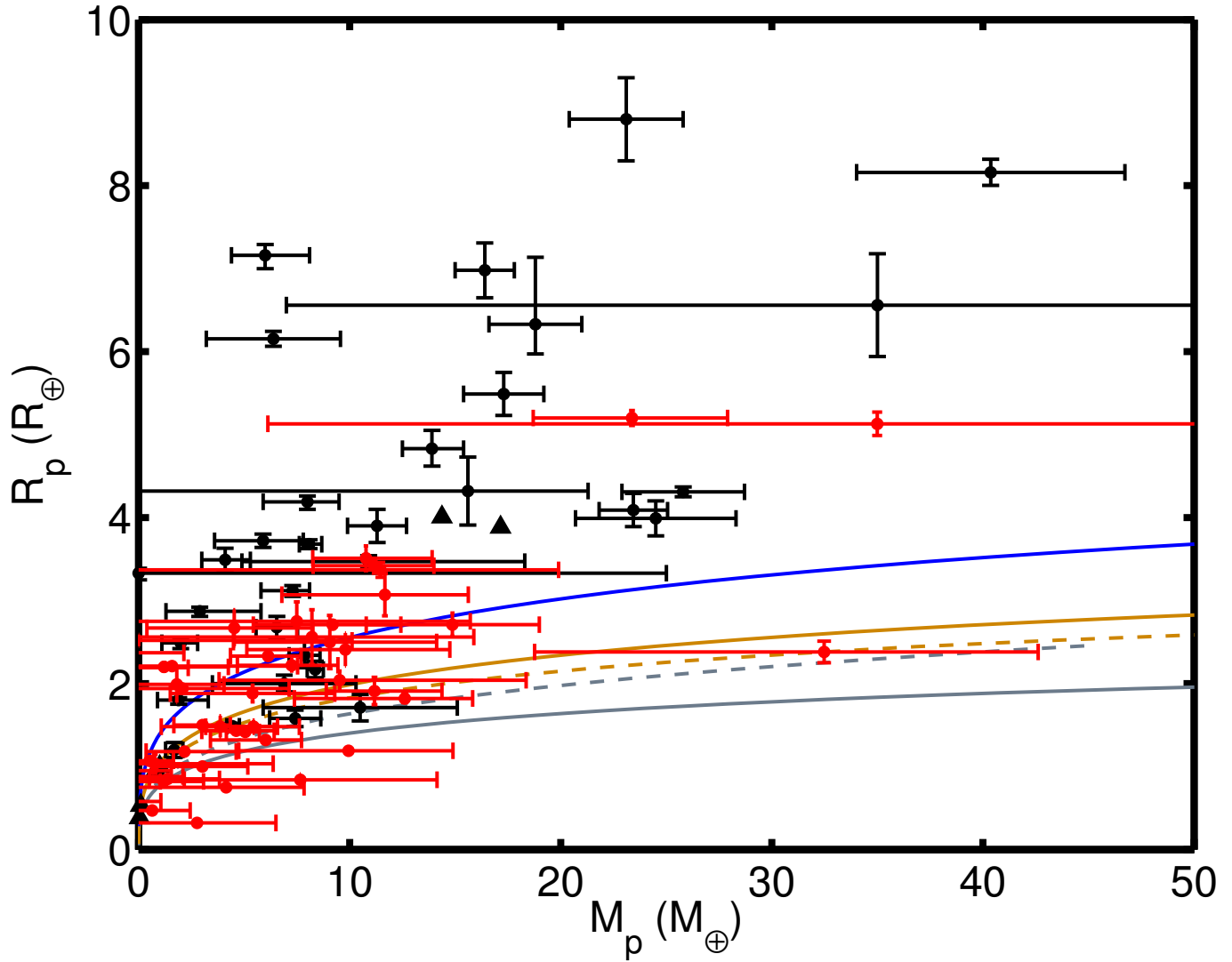}
\caption{Planet Mass Radius Diagram. The sample of {\it Kepler} planets with RV follow-up used in this work are highlighted in red. Other confirmed transiting sub-Neptune-size planets are indicated with black points, and the Solar System planets are indicated with black triangles. The colored curves are theoretical mass-radius relations for constant planet compositions from \citet{SeagerEt2007ApJ}: pure water ice (solid blue), pure MgSiO$_3$ silicate (solid brown), Earth-like composition (32\% Fe, 68\% silicate, dashed brown), maximum-density limit for rocky planets from simulations of collisional stripping \citep[][dashed gray]{MarcusEt2010ApJ} and pure Fe (solid gray).}
\label{fig:MRall} 
\end{figure}

One of the most direct insights we can glean about a sub-Neptune exoplanet's composition is whether it must have some volatiles (where volatiles refer to H/He or astrophysical ices). The mass-radius relation for a  silicate composition (the solid-brown curve in Figure~\ref{fig:MRall}) represents an extreme lower limit on the mass of rocky planet with no volatiles at a specified radius. Planets that are less dense must have some volatiles (in the form of water or H/He); though rock may still account for most of their mass, these planets are too low density to have their transit radius defined by a rocky surface. Planets that are more dense could potentially be comprised of iron and silicates alone. At the high-mass extreme, the mass-radius relation for a pure iron composition represents an upper limit to the mass of a rocky planet of a given size (the solid-gray curve in Figure~\ref{fig:MRall}). Planet mass-radius pairs that are more dense than pure iron are unphysical. 

We denote the minimum and maximum masses of rocky planets of a given radius, $R_p$, by $M_{\rm rock,min}\left(R_p\right)$ and $M_{\rm rock,max}\left(R_p\right)$, respectively. These mass-radius relations bound the ``potentially rocky" regime of planet mass-radius space. Planets with mass $M_{\rm rock,min}\left(R_p\right) \leq M_p\leq M_{\rm rock,max}\left(R_p\right)$ are potentially (but not necessarily) comprised of iron and silicates alone; planets in this regime may still contain substantial amounts of water and other volatiles if the low-density material is offset by higher density iron-enhanced rock. 

We nominally use the mass-radius relations from \citet{SeagerEt2007ApJ} for MgSiO$_3$ perovskite, and $\epsilon$ phase Fe to define $M_{\rm rock,min}\left(R_p\right)$ and $M_{\rm rock,max}\left(R_p\right)$. The curvature of these iso-composition mass-radius relations takes into account the compression of materials at higher pressure.  There is a maximum radius for planets in the ``potentially rocky" regime, set by the maximum radius of a sphere of the limiting low-density composition before degeneracy pressure takes over and causes radius to decrease with mass. For a pure silicate composition, $R_{\rm max, rock}=3.48~R_{\oplus}$. 

Pure-silicate and pure-iron are both hypothetical end-member compositions; more plausible limits to the masses of rocky planets would include some mixture of Fe and silicates at both the high and low mass extremes. By adopting a generous mass range in our definition of ``potentially rocky" planets at a given size, we set an upper bound on the fraction of planets that actually have a rocky composition. We investigate the effect of other choices for limiting high and low-density rocky planet mass-radius relations in Section~\ref{sec:EarthComp}. 

The lack of rocky planets with $R_p>2~R_{\oplus}$ (Figure~\ref{fig:MRall}) is one of the most striking (and largely model-independent) features of the transiting sub-Neptune-size planets with mass constraints. There is an apparent high-density/high-mass threshold to the measured planet masses and radii. The paucity of high-density/high-mass planets persists despite the fact that, for transit-discovered planets of a specified size, $R_p$, it is easier to detect denser and more massive planets in RV follow-up. This seems to indicate a dearth of rocky planets with masses in excess of $M_p\sim10~M_{\oplus}$. Our goal is to place this observation on a strong statistical footing. 

We must account for the large measurement uncertainties on planet masses and radii when assessing which planets may be rocky. Many of the {\it Kepler} planets with RV follow-up have either marginal RV detections or mass upper limits. Nonetheless, even the non-detections contain information about the planet composition. We introduce $p_{\rm rocky}$, the posterior probability that a planet is dense enough to be rocky based on the measured mass and radius. $p_{\rm rocky}$ is evaluated as the fraction of a transiting planet's joint mass-radius posterior probability density (obtained from fitting a planet model to the RV data and transit light curve as described in Section~\ref{sec:MRfit}) that falls within the high-density ``potentially rocky" regime ($M_{\rm rock,min}\left(R_p\right)\leq M_p\leq M_{\rm rock,max}\left(R_p\right)$). Planets with $p_{\rm rocky}$ near 1 are well constrained to have masses and radii in the potentially rocky regime, while low-density planets have $p_{\rm rocky}$ near 0. RV marginal detections or non-detections, may have measured RV amplitudes that spill into unphysical regimes (corresponding to negative mass or masses exceeding that of a pure iron sphere). In calculating $p_{\rm rocky}$, we assume a flat prior on planet mass-radius pairs that are physically plausible and a prior probability of 0 on mass radius pairs that are not (i.e., $p\left(M_p, R_p\right) = \rm{constant}$ if $M_p\leq M_{p{\rm rock,max}}\left(R_p\right)$ and $M_p>0$, otherwise $p\left(M_p, R_p\right) =0$). 

\begin{figure}
\centering
\includegraphics{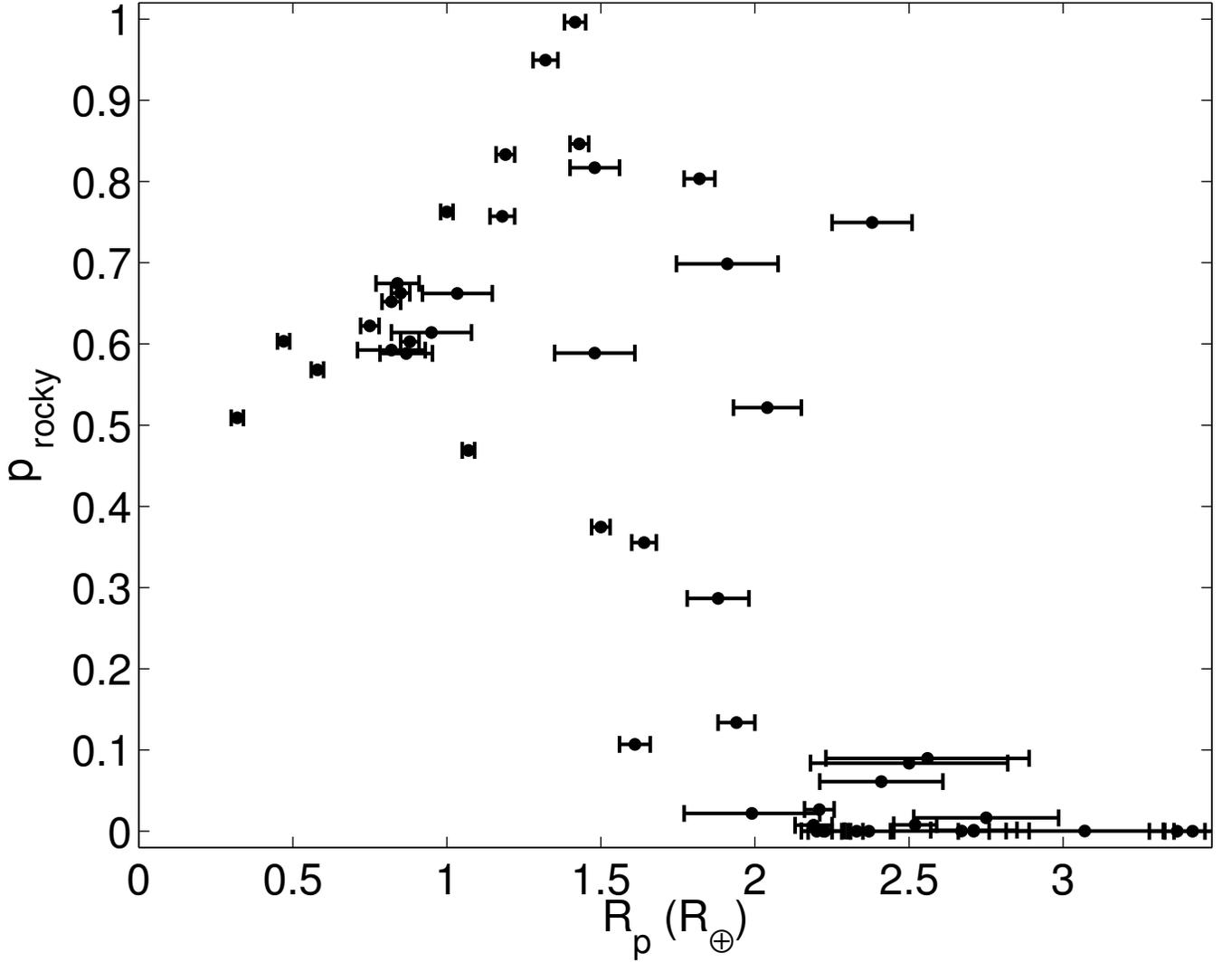}
\caption{Probability that a planet is sufficiently dense to be rocky, $p_{\rm rocky}$, as a function of planet size, $R_p$. The sample of {\it Kepler} transiting planets with Keck HIRES RV follow-up is plotted.}
\label{fig:RProcky} 
\end{figure}

The values of $p_{\rm rocky}$ for our sample of {\it Kepler} transiting planets with RV mass constraints are presented in Figure~\ref{fig:RProcky}. It is seen that smaller planets are more likely to be dense enough to be rocky. Planets larger than $\sim2~R_{\oplus}$ are too low density to be comprised of iron and silicates alone, there is some sort of transition regime between 1 and $2~R_{\oplus}$. 

\section{Statistical Methods for Characterizing Planet Population Properties}
\label{sec:stat}
While the trend of planets larger than $2~R_{\oplus}$ being volatile-rich is visible by eye in Figures~\ref{fig:MRall} and \ref{fig:RProcky}, we aim to quantify the rocky/non-rocky transition in a more statistically rigorous way. We would like to assess, (i) up to what size do the majority of planets have a rocky composition, and (ii) is the transition from rocky to non-rocky planet compositions as a function of $R_p$ gradual or abrupt. 

The calculation of $p_{\rm rocky}$ in the previous section (Figure~\ref{fig:RProcky}) considers each planet separately and independently, with a flat prior on each planet's mass and radius. We now show how a joint analysis of several planet systems having constraints on the planet masses and radii can be used to both constrain the mass-radius distribution of the planet population, and to inform our priors on the masses of subsequent transiting planet discoveries. 

We use a hierarchical Bayesian model to infer properties about the underlying population of planet compositions from the sample of {\it Kepler} planets with RV follow-up. The approach is hierarchical in the sense that we open up the priors assumed for each planet's mass and radius to modeling. Instead of assuming that any planet mass-radius pair is as likely or as unlikely as any other (the typical flat-prior assumption that has been applied in all planet mass-radius analyses to date), we assume the planets in our sample are drawn from a joint mass-radius distribution. We model the joint mass-radius distribution, assuming simple functional forms with a few free parameters, and derive the posterior probability density of those population-level parameters conditioned on the data. 

We describe our statistical approach in detail below. First, we set up a hierarchical Bayesian model framework to measure the underlying mass-radius distribution of a population of planets from a census of planets with constraints on their masses and radii (Section~\ref{sec:hbm}). Though we hope to eventually characterize the complete joint mass-radius distribution of planets, the dominant selection effects in the current sample of {\it Kepler} transiting planets with RV follow-up lead us to instead focus on the conditional distribution of planet mass at a specific planet radius. In Section~\ref{sec:MRprior}, we tailor the hierarchical Bayesian approach to constrain the fraction of planets, $f_{\rm rocky}\left(R_p\right)$ that are sufficiently dense to be rocky, as a function of planet size. 

\subsection{Hierarchical Bayesian Model for the Planet Mass--Radius Distribution}
\label{sec:hbm}

We adapt the resampling approach outlined by \citet{HoggEt2010ApJ}, which focussed on constraining the true distribution of planet eccentricities from the likelihood functions on the orbital parameters of RV-detected planets, to the question of constraining the underlying true joint distribution of planet masses and radii from a sample of transiting planets with RV follow-up. This method takes as input a sample from the posterior probability density for the masses and radii in the individual planetary systems. 

We have $N$ stars (indexed by $n$, where $1\leq n\leq N$) that are orbited by at least one transiting planet, and that have each been followed up with spectroscopic RV measurements. Each star is orbited by $J_n$ planets detected in transit (indexed by $j$,  where $1\leq j\leq J_n$). We use $\boldsymbol{D}_n$  to denote the data on the $n$th star, including RV measurements and the light curve photometry.  

We denote by $\boldsymbol{\beta}_n$ the properties of the planet(s) orbiting the $n$th star. In this work, we will focus mostly on planet masses $M_{nj}$ and radii $R_{nj}$, but $\boldsymbol{\beta}_n$ also includes the number of planets orbiting the star, the planet orbital periods, host star properties, and other model parameters from the photodynamical fit that, for this work, are nuisance parameters that we marginalize over (e.g., star center of mass velocity, transit ephemeris, properties of non-transiting planets detected by RV-trends etc).  

For each star, $n$, we imagine that we have been provided \citep[in our case by ][]{MarcyEt2014ApJS} with a $K_n$-element sampling from the posterior probability density function (PDF) of the planet properties obtained from fitting the RV and photometry data,
\begin{equation}
p\left(\boldsymbol{\beta}_n\left|\right.\boldsymbol{D}_n\right)= \frac{1}{Z_n}p\left(\boldsymbol{D}_n\left|\right.\boldsymbol{\beta}_n\right)p_0\left(\boldsymbol{\beta}_n\right)
\end{equation}
\noindent Above, $\mathscr{L}_n=p\left(\boldsymbol{D}_n\left|\right.\boldsymbol{\beta}_n\right)$ is the likelihood of the data given a set of planet model parameters $\boldsymbol{\beta}_n$, $Z_n$ is a normalization constant, and $p_0\left(\boldsymbol{\beta}_n\right)$ is an uninformative ``interim" prior PDF chosen by the RV-photometry data fitter \citep[][took a flat prior on planet mass and radius]{MarcyEt2014ApJS}. For each planetary system, $n$, the sampling takes the form of a chain of $K_n$ samples (indexed by $k$, where $1\leq k\leq K_n$), each a complete set of the planet parameters $\beta_{nk}$, such that the distribution of samples is consistent with a random draw from the posterior PDF, $p\left(\boldsymbol{\beta}_n\left|\right.\boldsymbol{D}_n\right)$. 

Assuming that the observations of different stars are independent (i.e., neglecting any correlations between observations of different stars introduced by hardware issues or calibration), the total posterior, likelihood and prior for all the parameters of all the $N$ planetary systems, is just the product of the individual system posteriors, likelihoods and priors, respectively 
\begin{eqnarray}
p\left(\left\{\boldsymbol{\beta}_n\right\}_{n=1}^N\left|\right.\left\{\boldsymbol{D}_n\right\}_{n=1}^N\right)&=&\prod_{n=1}^{N}p\left(\boldsymbol{\beta}_n\left|\right.\boldsymbol{D}_n\right) \label{eq:totpost}\\
\mathscr{L}\equiv p\left(\left\{\boldsymbol{D}_n\right\}_{n=1}^N\left|\right.\left\{\boldsymbol{\beta}_n\right\}_{n=1}^N\right)&=&\prod_{n=1}^{N}\mathscr{L}_n \label{eq:totlik}\\
p_0\left(\left\{\boldsymbol{\beta}_n\right\}_{n=1}^N\right)&=&\prod_{n=1}^{N}p_0\left(\boldsymbol{\beta}_n\right). \label{eq:totprior}
\end{eqnarray}

Our goal is to constrain the distribution of ``true" (as opposed to measured or estimated) planet masses and radii based on the noisy observations of all $N$ planetary systems. Here, the adjective ``true" refers to what would have been measured in \emph{much} higher signal-to-noise ratio observations of the same objects. To this end, we develop a model for the distribution of true planet masses and radii, that depends on some population-level parameters $\boldsymbol{\alpha}$. We use that population model to define a new set of priors on the planet properties, depending on $\boldsymbol{\alpha}$
\begin{equation}
p\left(\left\{\boldsymbol{\beta}_n\right\}_{n=1}^N\left|\right.\boldsymbol{\alpha}\right) = \prod_{n=1}^{N}p\left(\boldsymbol{\beta}_n\left|\right.\boldsymbol{\alpha}\right) 
\end{equation}
\noindent The joint posterior probability for the properties of all the individual planets and the population-level parameters $\boldsymbol{\alpha}$ governing the planet mass-radius distribution is then,
\begin{equation}
p\left(\left\{\boldsymbol{\beta}_n\right\}_{n=1}^N, \boldsymbol{\alpha}\left|\right.\left\{\boldsymbol{D}_n\right\}_{n=1}^N\right)\propto p\left(\left\{\boldsymbol{D}_n\right\}_{n=1}^N\left|\right.\left\{\boldsymbol{\beta}_n\right\}_{n=1}^N\right)p\left(\left\{\boldsymbol{\beta}_n\right\}_{n=1}^N\left|\right.\boldsymbol{\alpha}\right)p\left(\boldsymbol{\alpha}\right),\label{eq:postall}
\end{equation}
\noindent where $p\left(\boldsymbol{\alpha}\right)$ are the priors on the population-level parameters. 

The posterior presented above in Equation~(\ref{eq:postall}) is hierarchical: the parameters describing the distribution of planet properties are inferred, and influence the estimates of the individual planet properties. With this hierarchical framework, characterizing the true planet mass-radius distribution boils down to constraining the population-level parameters, $\boldsymbol{\alpha}$. For this purpose, the true properties of the individual planetary systems are nuisance parameters. Marginalizing over (integrating out) the nuisance parameters, we obtain the posterior PDF on $\boldsymbol{\alpha}$, 
\begin{equation}
p\left(\boldsymbol{\alpha}\left|\right.\left\{\boldsymbol{D}_n\right\}_{n=1}^N\right)\propto \left[ \prod_{n=1}^{N} \int d\beta_n p\left(\boldsymbol{D}_n\left|\right.\boldsymbol{\beta}_n\right)p\left(\boldsymbol{\beta}_n\left|\right.\boldsymbol{\alpha}\right)\right]p\left(\boldsymbol{\alpha}\right).
\end{equation}
\noindent We may then identify the marginal likelihood function of  $\boldsymbol{\alpha}$, $\mathscr{L}_{\boldsymbol{\alpha}}$,
\begin{equation}
\mathscr{L}_{\boldsymbol{\alpha}}\equiv p\left(\left\{\boldsymbol{D}_n\right\}_{n=1}^N\left|\right.\boldsymbol{\alpha}\right)= \prod_{n=1}^{N} \int d\beta_n p\left(\boldsymbol{D}_n\left|\right.\boldsymbol{\beta}_n\right)p\left(\boldsymbol{\beta}_n\left|\right.\boldsymbol{\alpha}\right).\label{eq:likealpha1}
\end{equation}

Fortunately, we do not have to evaluate the multi-dimension integrals in Equation~(\ref{eq:likealpha1}) directly. The marginalized likelihood for $\boldsymbol{\alpha}$ can readily be evaluated by applying importance resampling to the samples from the posterior PDF for each individual planet system's properties provided to us by the RV-photometry fitter. 
\begin{equation}
\mathscr{L}_{\boldsymbol{\alpha}}\approx  \prod_{n=1}^{N} \frac{1}{K_n}\sum_{k=1}^{K_n} \frac{p\left(\boldsymbol{\beta}_{nk}\left|\right.\boldsymbol{\alpha}\right)}{p_0\left(\boldsymbol{\beta}_{nk}\right)} \label{eq:Laresample}
\end{equation}
\noindent Equation~(\ref{eq:Laresample}) above is equivalent to Equation~(9) of \citet{HoggEt2010ApJ}. Each element in the PDF samples for an individual planet system is re-weighted by the ratio of the new prior PDF (depending on $\boldsymbol{\alpha}$, which we want to infer) and the interim prior PDF (on which the original sampling was based). We elaborate upon the choice of $p\left(\boldsymbol{\beta}_{nk}\left|\right.\boldsymbol{\alpha}\right)$ in the next section.

\subsection{Modeling the Conditional Planet Mass Distribution, at Specified Planet Radius}
\label{sec:MRprior}

In the previous section we described a general approach to constraining the true mass-radius distribution of planets from a census of transiting planets with RV follow-up. We now tailor the approach to the \citet{MarcyEt2014ApJS} sample of {\it Kepler} planets with RV follow-up. 

As described in Section~\ref{sec:statMR}, the \citet{MarcyEt2014ApJS} sample of sub-Neptune-size planets were initially detected from transits in the {\it Kepler} photometry, selected for RV follow-up based on their radius, orbital period and stellar properties, and then spectroscopically followed-up with Keck-HIRES in a mass-blind way. 
For a specified planet radius, the measured RV-masses of the \citet{MarcyEt2014ApJS}  planets provide an unbiased sampling of close-in planet masses of that particular size. In contrast, the radius distribution of the \citet{MarcyEt2014ApJS} planets is dominated by the selection effects applied in choosing {\it Kepler} planet candidates, and does not reflect the true underlying radius distribution of planets. 

In our hierarchical Bayesian analysis, we factor out the dominant selection effects by focussing on the conditional distribution of planet masses at a specified planet radius. We frame our model for the conditional distribution of planet masses at a specified radius in terms of the fraction of planets of that size that are sufficiently dense to be rocky, $f_{\rm rocky}\left|\right.R_{p}$. As defined, $f_{\rm rocky}\left|\right.R_{p}$ is the fraction of planets of radius $R_p$ that have mass greater than $M_{\rm rock,min}\left(R_p\right)$, while  $1-f_{\rm rocky}\left|\right.R_{p}$ is the fraction of planets with radius $R_p$ that have mass less than $M_{\rm rock,min}\left(R_p\right)$ (and thus must contain astrophysical ices or H/He that contribute to the transit depth). 

We assume simple functional forms for $f_{\rm rocky}\left|\right.R_{p}$, that depend on a few free parameters $\boldsymbol{\alpha}$, the planet radius, and (potentially) other properties of the planet-star system (such as planet insolation or stellar mass)
\begin{equation}
f_{\rm rocky}\left|\right.R_{p}= f_{\alpha}\left(R_{p}\right).
\end{equation}
\noindent In our model, the distribution of planet masses conditioned on a particular planet radius depends on $f_{\rm rocky}\left|\right.R_{p}$ (and hence on $\boldsymbol{\alpha}$). Instead of a completely flat prior on planet mass (as typically assumed when fitting individual planet systems), we divide mass into two regimes (``potentially rocky" and ``non-rocky"), and take a flat prior PDF on the planet mass within each regime
\begin{equation}
p\left(M_{nj}\left|\right.R_{nj}, \boldsymbol{\alpha}\right) = \left\{
     \begin{array}{ll}
       \frac{1-f_{\alpha}\left(R_{nj}\right)}{ M_{p{\rm rock,min}}} &  M_{nj} <  M_{p{\rm rock,min}}\left(R_{nj}\right)\\
        \frac{f_{\alpha}\left(R_{nj}\right)}{ M_{p{\rm rock,max}}-M_{p{\rm rock,min}}} & M_{p{\rm rock,min}}\left(R_{nj}\right) \leq M_{nj} \leq M_{p{\rm rock,max}}\left(R_{nj}\right)\\
       0 &  \rm{otherwise}.
     \end{array}
   \right.
   \label{eq:flatm}
\end{equation} 
\noindent On all other parameters except for planet mass, $M_p$, we keep the same non-informative interim prior PDF as used by \citet{MarcyEt2014ApJS}. Assuming separability of the prior PDF, 
\begin{equation}
p\left(\boldsymbol{\beta}_{n}\left|\right.\boldsymbol{\alpha}\right) = \left[\prod_{j=1}^{J_n}\frac{p\left(M_{nj}\left|\right. R_{nj}, \boldsymbol{\alpha}\right)}{p_0\left(M_{nj}\right)}\right]p_0\left(\boldsymbol{\beta}_{n}\right).\label{eq:prioralpha}
\end{equation}
\noindent The interior prior, $p_0\left(M_{nj}\right)$, is flat on $M_p$ for each planet $j$ in the $n$th system. The $\alpha$-dependent prior we hope to infer also treats each planet independently.

Using Equations~(\ref{eq:Laresample}) and (\ref{eq:prioralpha}) we constrain the posterior PDF of $\boldsymbol{\alpha}$. Then, based on the assumed functional form for $f_{\alpha}\left(R_{p}\right)$, we may transform $p\left(\boldsymbol{\alpha}\left|\right.\left\{\boldsymbol{D}_n\right\}_{n=1}^N\right)$ to obtain a posterior PDF for $f_{\rm rocky}$, conditioned on planet size, $R_p$. 
\begin{equation}
p\left(f_{\rm rocky} \left|\right. R_p,  \left\{\boldsymbol{D}_n\right\}_{n=1}^N\right)=\int p\left(\boldsymbol{\alpha}\left|\right.\left\{\boldsymbol{D}_n\right\}_{n=1}^N\right)\delta\left(f_{\alpha}\left(R_p, \boldsymbol{\alpha}\right)-f_{\rm rocky}\right)d\boldsymbol{\alpha},\label{eqn:frockyposterior}
\end{equation}
\noindent where $\delta$ is the Dirac-delta function. The posterior PDF $p\left(f_{\rm rocky} \left|\right. R_p, \left\{\boldsymbol{D}_n\right\}_{n=1}^N\right)$ quantifies and summarizes the constraints the {\it Kepler} planet candidates with RV follow-up place on the fraction $f_{\rm rocky}$ of planets that are dense enough to be rocky as a function of planet size. For a given $R_p$, values of $f_{\rm rocky}$ with higher values of $p\left(f_{\rm rocky} \left|\right. R_p,  \left\{\boldsymbol{D}_n\right\}_{n=1}^N\right)$ are more strongly favored by the RV and photometry data. This approach to constraining the fraction of planets of a given size that are rocky takes into account both the observational uncertainties on the individual planet masses and radii, and the statistical uncertainties due to the finite number of planets in our sample.

\section{Results}
\label{sec:res}

We now apply the hierarchical Bayesian formalism described above to the \citet{MarcyEt2014ApJS} sample of transiting planets with mass constraints. We explore several different choices for the functional form of $f_{\alpha}\left(R_{p}\right)$.

\subsection{Step-function Rocky/Non-rocky Radius Threshold}

We first explore the possibility that all planets larger than a threshold radius, $R_{\rm thresh}$, are non-rocky, while all planets smaller than $R_{\rm thresh}$ are dense enough to be comprised of iron and silicates alone. In this scenario, we have a one-parameter ($\boldsymbol{\alpha}_1 \equiv R_{\rm thresh}$) step-function model where the fraction of planets with volatiles depends only on planet radius, $f_{1\alpha}\left(R_p, R_{\rm thresh}\right)$, 
\begin{equation}
f_{1\alpha}\left(R_p, R_{\rm thresh}\right) = \begin{cases}
        1 &  R_p < R_{\rm thresh}\\   
        0 &  R_p\geq R_{\rm thresh}\\
       \end{cases}.
         \label{eq:fvol1}
\end{equation}
\noindent We adopt uniform priors on $0<R_{\rm thresh}<R_{\rm max, rock}$, where $R_{\rm max, rock}=3.48~R_{\oplus}$ is a very conservative upper bound on the radius of rocky planets (the maximum radius of a silicate sphere before degeneracy pressure takes over and causes radius to decrease with mass). This simple one-parameter model serves as a baseline case against which all more sophisticated models are compared.

\begin{figure}
\centering
\includegraphics{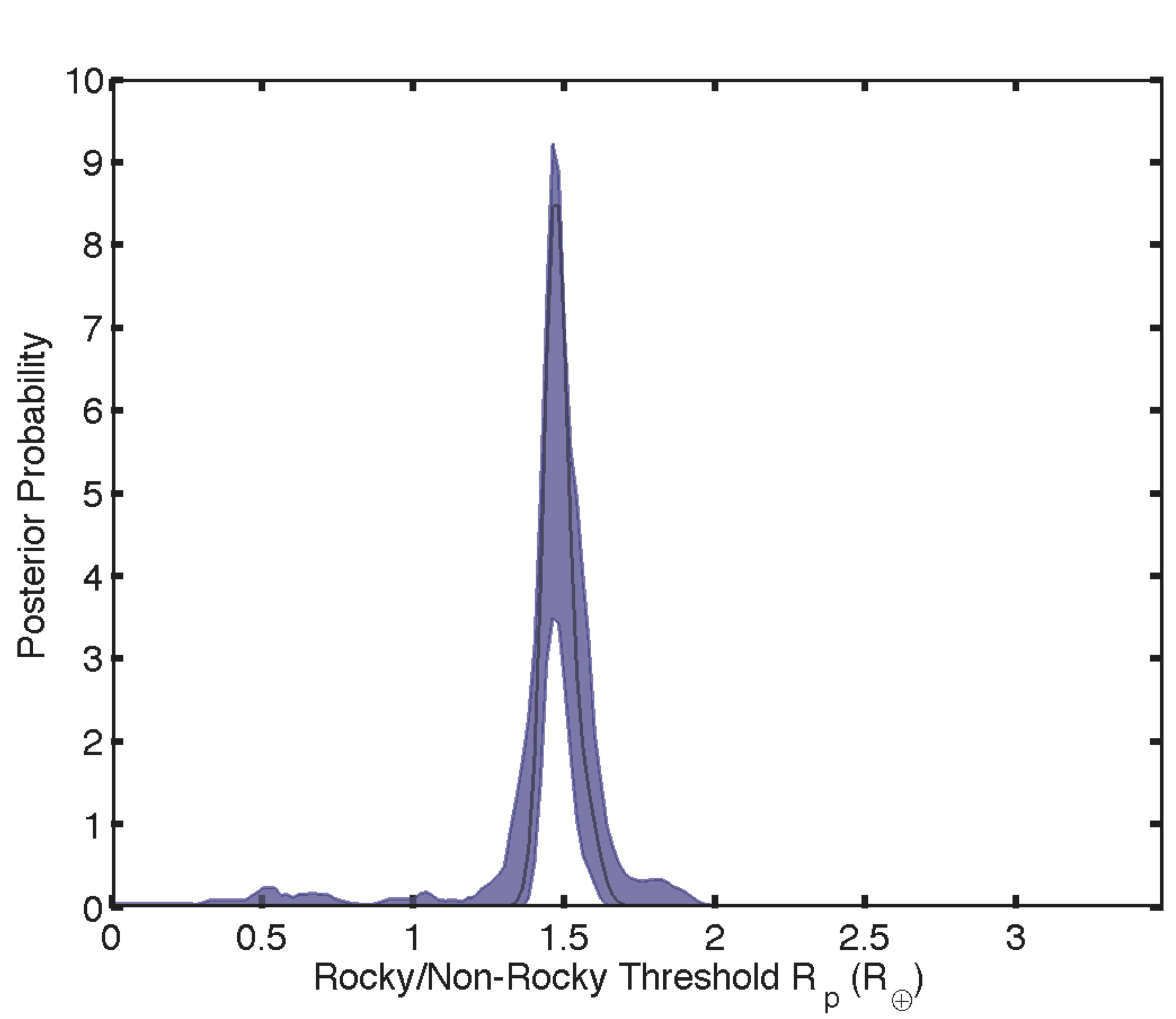}
\caption{Posterior PDF for $\boldsymbol{\alpha}_1 \equiv R_{\rm thresh}$, in the one-parameter step-function model for $f_{\rm rocky}\left|\right.R_p$ (Equation~(\ref{eq:fvol1})). The black curve gives the posterior probability of the rocky/non-rocky threshold, wherein all planet larger than $R_{\rm thresh}$ have volatiles, while all planets smaller than $R_{\rm thresh}$ are dense enough to be rocky. The blue shaded uncertainty regions encompass 68.3\% of the values of $p\left(R_{\rm thresh}\left|\right. \rm{data}\right)$ evaluated from 1000 bootstrapping samples.}
\label{fig:fvol1} 
\end{figure}

Figure~\ref{fig:fvol1} presents the resulting posterior probability on $R_{\rm thresh}$. We find a median value $R_{\rm thresh} = 1.48~R_{\oplus}$, and a 95\% confidence upper bound of $R_{\rm thresh} = 1.59~R_{\oplus}$. The finding that $R_{\rm thresh}$ is near $1.5~R_{\oplus}$ with a 95\% confidence upper limit of $1.6~R_{\oplus}$  effectively quantifies the trend visible by eye in Figure~\ref{fig:MRall} that most planets discovered to date that are larger than $1.6~R_{\oplus}$ are too low-density to be rocky. For the pure silicate composition adopted as the low-density extreme for rocky planets, the median and 95\% upper bound on $R_{\rm thresh}$ correspond to threshold masses of $3.5~M_{\oplus}$ and $5.0~M_{\oplus}$, respectively. For an Earth-like composition, the median and 95\% upper bound on  $R_{\rm thresh}$ correspond to masses of $4.5~M_{\oplus}$ and $6.0~M_{\oplus}$, respectively. 

\subsection{Gradual Rocky/Non-rocky Radius Threshold}

Physically, we might expect that small planets will show some composition diversity, and that there will be a range of radii at which high-mass rocky planets and low-mass non-rocky planets co-exist. We now investigate whether there is evidence in the data for a gradual transition between the populations of rocky and non-rocky planets, and how allowing the possibility of a gradual transition affects the inferred relative frequency of rocky planets at a given size.  

\begin{figure}
\centering
\includegraphics[width=0.75\textwidth]{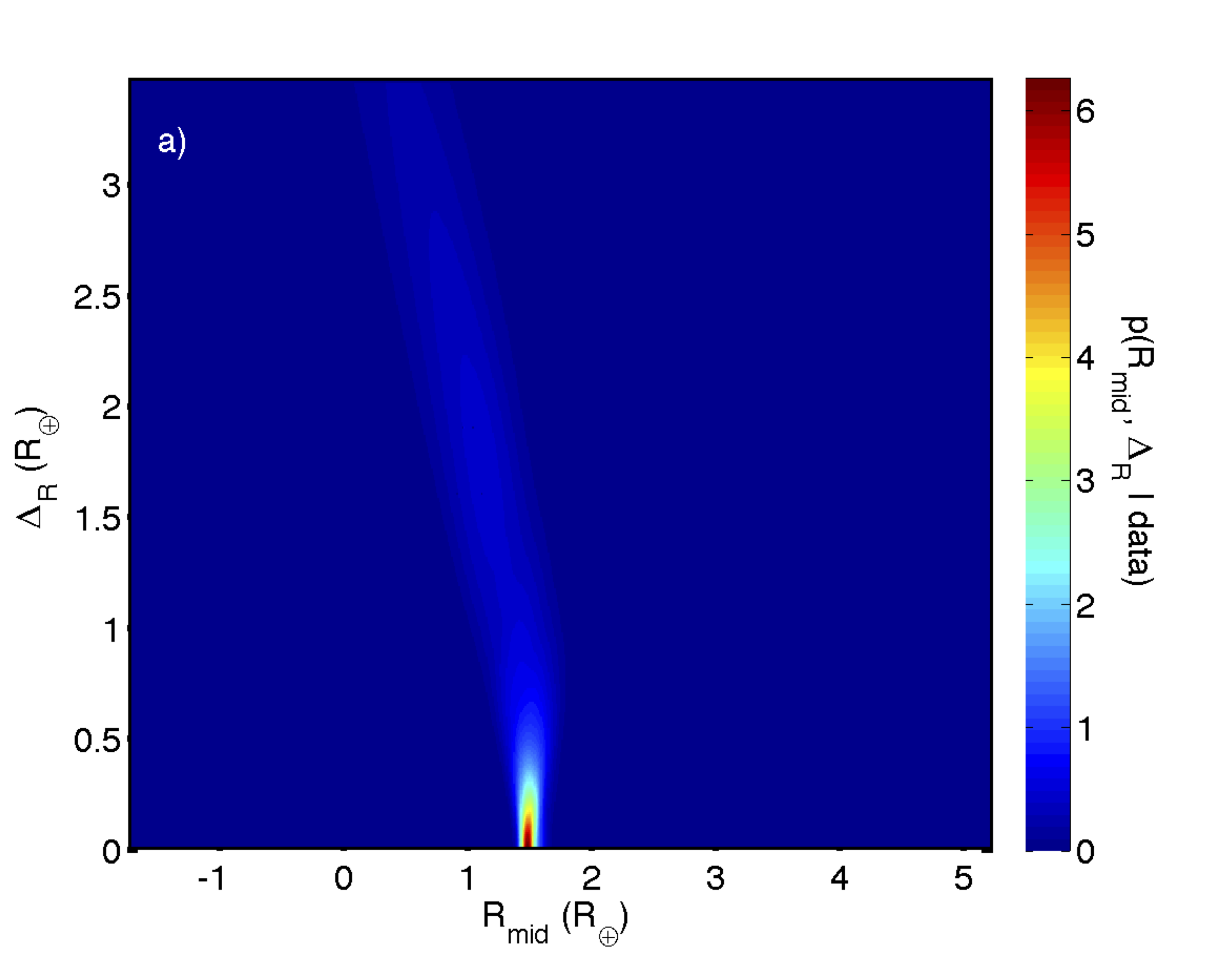}
\includegraphics[width=0.75\textwidth]{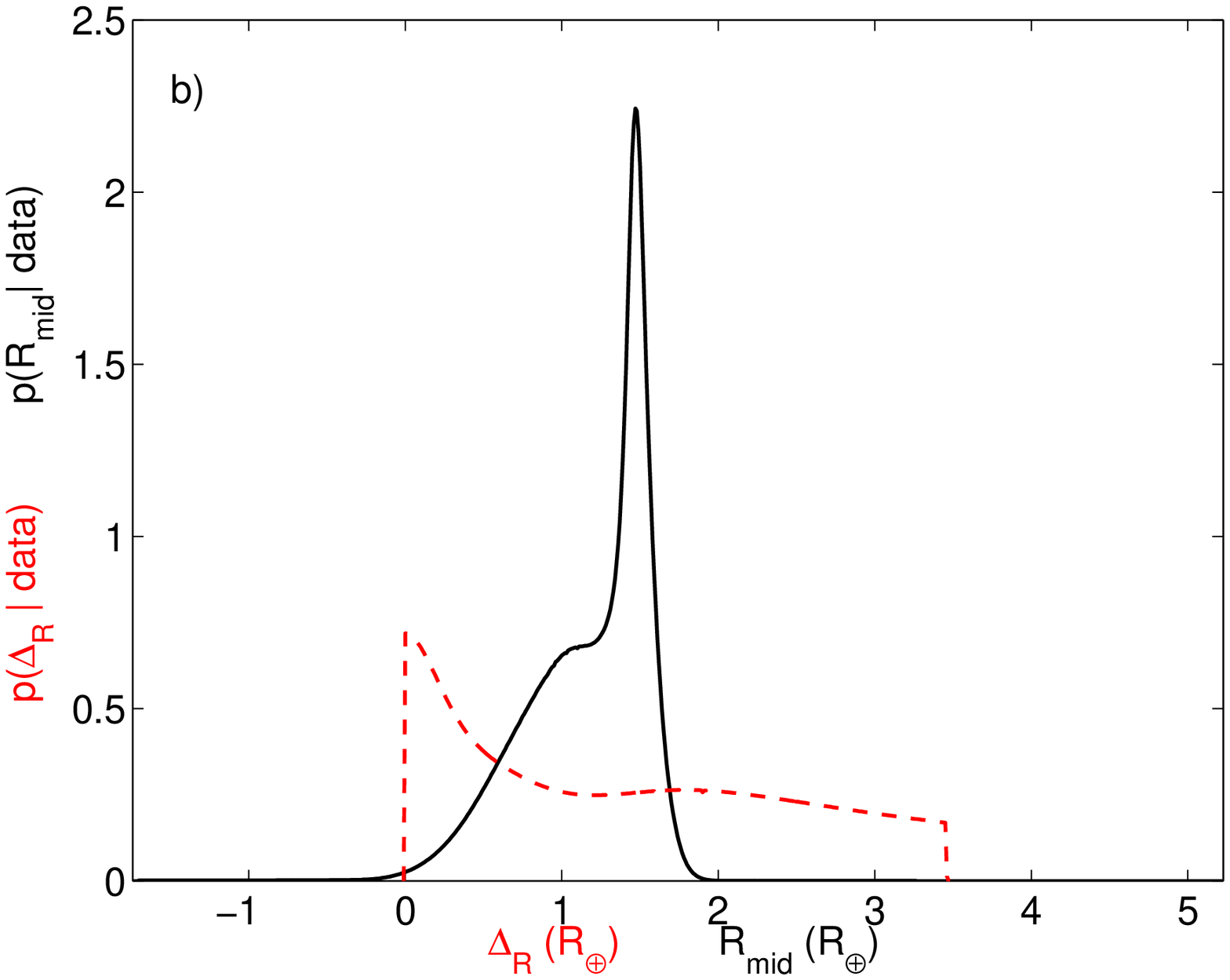}
\caption{Posterior PDF for $\boldsymbol{\alpha}_2  \equiv \left\{ R_{\rm mid}, \Delta_R \right\}$ in the linear rocky/non-rocky transition model for $f_{\rm rocky}\left|\right.R_p$ (Equation~(\ref{eq:fvol2})). The upper panel (a) shows the joint posterior  distribution on $\left(R_{\rm mid}, \Delta_R\right)$, while the bottom panel (b) shows the marginalized distribution of each parameter.  }
\label{fig:fvol2} 
\end{figure}

In our second model, assume that the fraction of planets that are dense enough to be rocky, $f_{\rm rocky}$ decreases with planet radius in a piecewise-linear fashion 
\begin{equation}
f_{2\alpha}\left(R_p, R_{\rm mid}, \Delta_R\right) = \begin{cases}
 	1 &  R_p\leq R_{\rm mid} -\frac{1}{2}\Delta_R\\
       0.5 + \frac{R_{\rm mid}-R_p}{\Delta_R} &  R_{\rm mid} -\frac{1}{2}\Delta_R<R_p<R_{\rm mid} + \frac{1}{2}\Delta_R\\
       0 &  R_p\geq \rm{min}\left(R_{\rm mid} + \frac{1}{2}\Delta_R, R_{\rm max, rock}\right).\\
       \end{cases}
         \label{eq:fvol2}
\end{equation}
\noindent We are still assuming that the non-rocky planet fraction depends only on planet radius. However, our model now has two parameters ($\boldsymbol{\alpha}_2 \equiv \left\{ R_{\rm mid}, \Delta_R \right\}$),  the mid-point of the linear transition (where $f_{2\alpha}=0.5$), $R_{\rm mid}$, and the width of the transition, $\Delta_R$. This linear transition model  reduces to the step-function model when $\Delta_R=0$, but admits the possibility of a range of radii over which rocky and non-rocky planets coexist. For our priors, we take a flat prior distribution on $\left(R_{\rm mid}, \Delta_R \right)$ from the rectangular 2D area, $-0.5R_{\rm max, rock}<R_{\rm mid}<1.5R_{\rm max, rock}$, $0<\Delta_R<R_{\rm max, rock}$. 
In cases with $R_{\rm mid} -\frac{1}{2}\Delta_R<0$ there is a finite prior probability of non-rocky planets as $R_p$ approaches 0, and when $R_{\rm mid} + \frac{1}{2}\Delta_R>R_{\rm max, rock}$ there is a finite prior probability of rocky planets up to the maximum physically allowable size, $R_{\rm max, rock}$. The joint posterior probability density of $R_{\rm mid}$ and $\Delta_R$ conditioned on the observed planets is displayed in Figure~\ref{fig:fvol2}, along with the marginal distribution of each parameter. 

At any radius equal to or larger than $1.62~R_{\oplus}$, the majority (50\% or more) of planets of that size are not-rocky, (at 95\% statistical confidence, based on the linear transition model for $f_{\rm rocky}$). In the linear transition model, $R_{\rm mid}$ represents the planet size at which 50\% of the planets are sufficiently dense to be rocky. Marginalizing over $\Delta_R$ (Figure~\ref{fig:fvol2}), we constrain $R_{\rm mid}=1.29^{+0.23}_{-0.54}~R_{\oplus}$, where we list the median as the nominal value and quote the $34.1\%$ credible interval on either side of the median as the uncertainties. The maximum probability (mode) is $R_{\rm mid}=1.47~R_{\oplus}$, and $1.62~R_{\oplus}$ is the 95th percentile. Iso-probability contours in $R_{\rm mid}$--$\Delta R$ space encompassing 68.27, 95.45 and 99.73\% of the posterior probability include scenarios with $R_{\rm mid}$ up to $1.70~R_{\oplus}$, $1.79~R_{\oplus}$ and $1.87~R_{\oplus}$, respectively. 

 The width $\Delta_R$ of the transition between rocky and non-rocky planets is consistent with an abrupt transition $\left(\Delta_R=0\right)$. The mode of the marginalized posterior distribution is indeed found at $\Delta_R=0$. The distribution of $\Delta_R$ is very broad, with a median and $1\sigma$ confidence interval of $\Delta_R=1.28^{+1.34}_{-1.04}~R_{\oplus}$. Notably, the maximum probability solution of the linear-transition model  ($R_{\rm mid}=1.47~R_{\oplus}$,  $\Delta R=0.02~R_{\oplus}$) nearly coincides with the best fit of the step-function model.

With the added flexibility of this 2-parameter linear transition model, the 68.27\% ($1~\sigma$ ) and 95.45\% ($2~\sigma$) credible regions include solutions with rocky planets extending to larger radii as compared to the step-function model. In the linear transition model, the maximum radius achieved by rocky planets (the smallest radius at which  $f_{2\alpha}=0$) is $R_{\rm mid} + \frac{1}{2}\Delta_R$. Iso-probability contours in $R_{\rm mid}$--$\Delta R$ space encompassing 68.27, 95.45 and 99.73\% of the posterior probability include values of $R_{\rm mid} + \frac{1}{2}\Delta_R\leq 2.27~R_{\oplus}$, $2.64~R_{\oplus}$ and $2.95~R_{\oplus}$, respectively. We similarly set upper limits on the planet radius at which no more than 5\% of planets are dense enough to be rocky of $2.15~R_{\oplus}$ (68.27\%), $2.46~R_{\oplus}$ (95.45\%), and $2.78~R_{\oplus}$ (99.73\%).  

\begin{figure}
\centering
\includegraphics[width=0.75\textwidth]{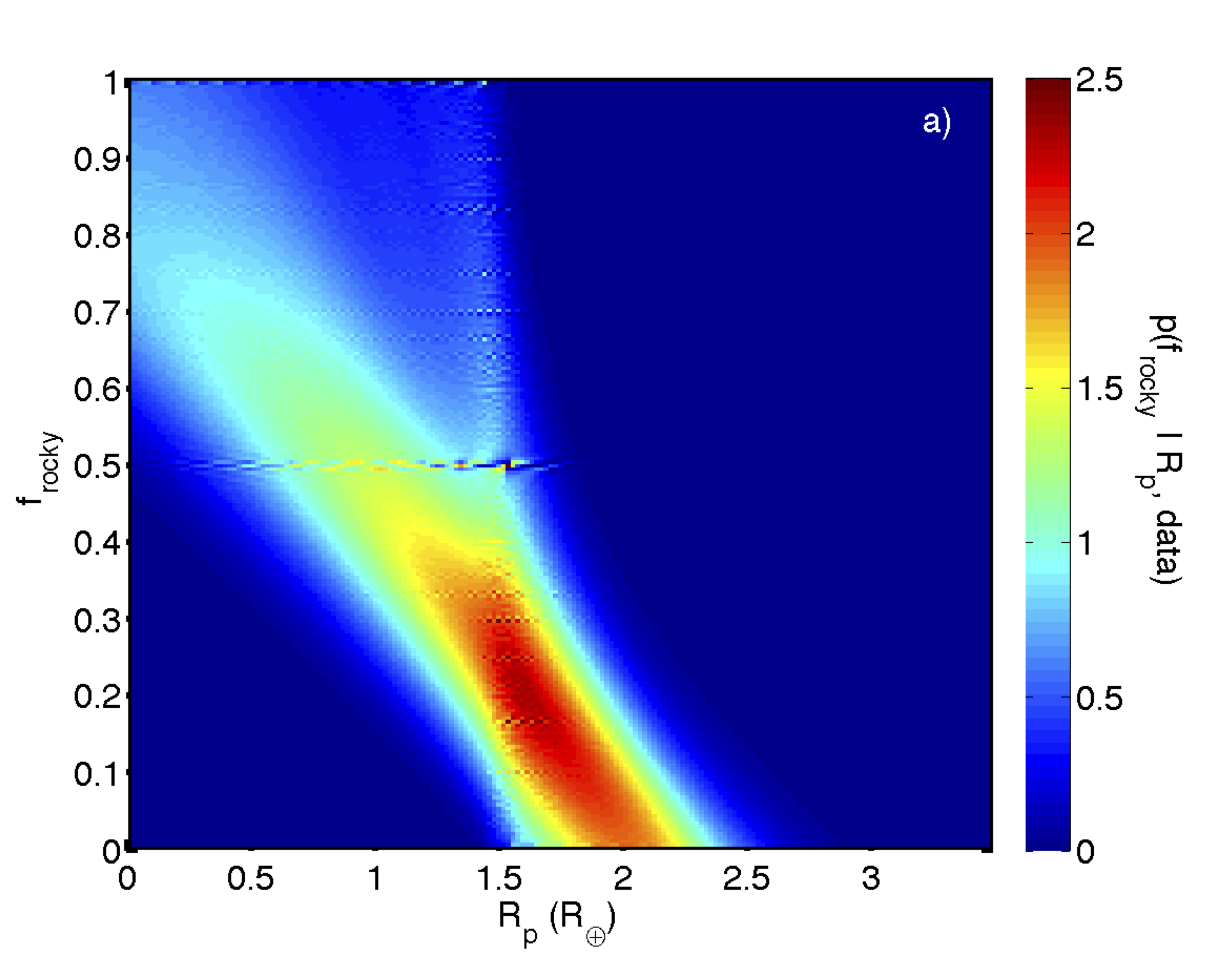}
\includegraphics[width=0.75\textwidth]{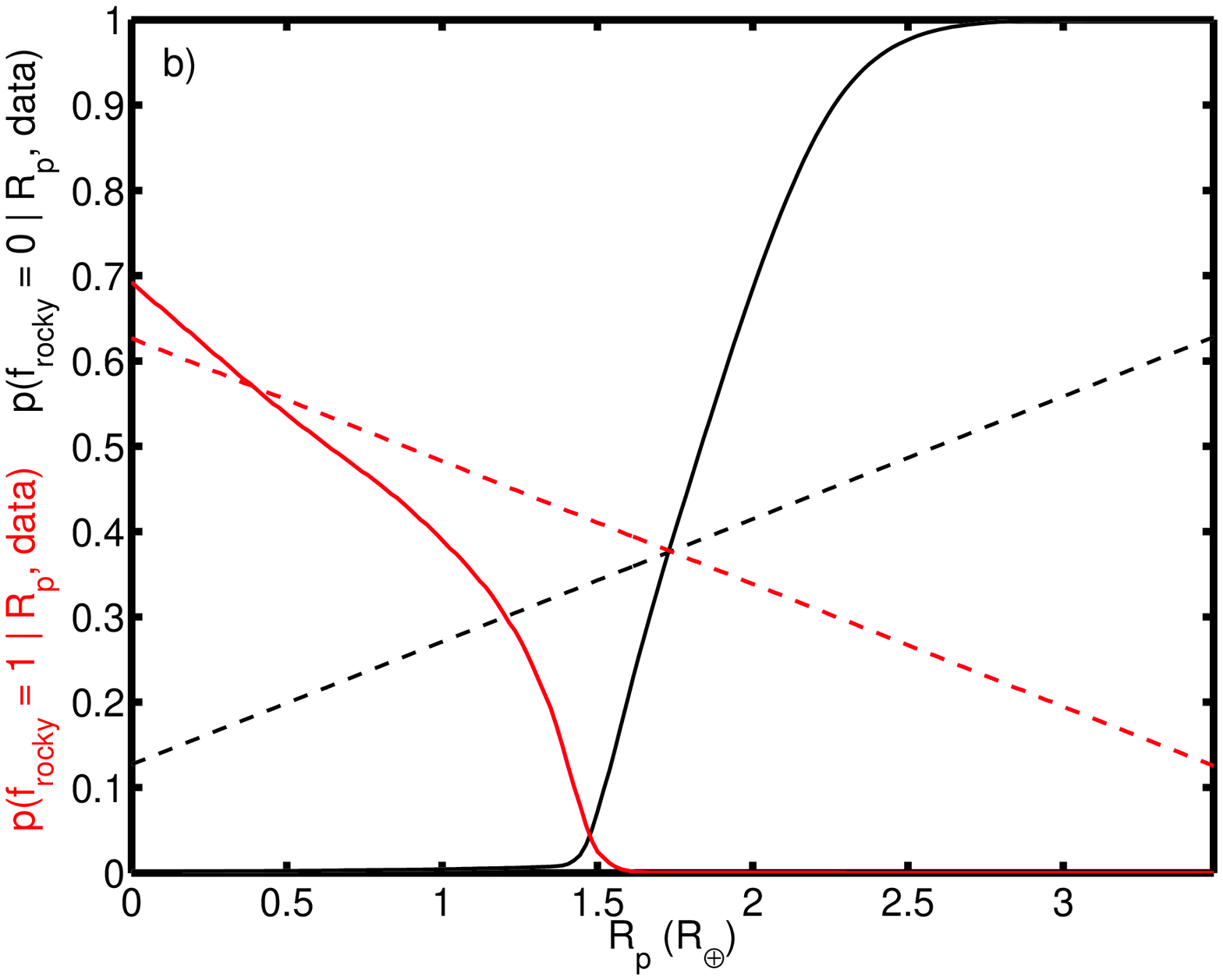}
\caption{Posterior probability density $p\left(f_{\rm rocky} \left|\right. R_p, \left\{\boldsymbol{D}_n\right\}_{n=1}^N\right)$ on the fraction of planets that are sufficiently dense to be rocky, as a function of planet size, obtained by applying the linear transition model (Equation~(\ref{eq:fvol2})). The posterior PDF of $f_{\rm rocky}$ is discontinuous at $f_{\rm rocky}=0$ and $f_{\rm rocky}=1$; the top panel (a) presents $p\left(f_{\rm rocky} \left|\right. R_p, \left\{\boldsymbol{D}_n\right\}_{n=1}^N\right)$ for $0<f_{\rm rocky}<1$, while the bottom panel (b) presents $p\left(f_{\rm rocky}=0 \left|\right. R_p, \left\{\boldsymbol{D}_n\right\}_{n=1}^N\right)$ (solid black line) and $p\left(f_{\rm rocky}=1 \left|\right. R_p, \left\{\boldsymbol{D}_n\right\}_{n=1}^N\right)$ (solid red line).  The value of the priors, $p_0\left( f_{\rm rocky}\left|\right. R_p\right)$, at $f_{\rm rocky}=0$ and $f_{\rm rocky}=1$ are shown with the dashed black and red lines, respectively.}
\label{fig:frockycondRp2} 
\end{figure}

The posterior PDF of $f_{\rm rocky}$ conditioned on planet radius, $p\left(f_{\rm rocky} \left|\right. R_p, \left\{\boldsymbol{D}_n\right\}_{n=1}^N \right)$,  quantifies the constraints that the {\it Kepler} planet candidates with Keck RV follow-up place on the fraction $f_{\rm rocky}$ of planets that are dense enough to be rocky as a function of planet size (Figure~\ref{fig:frockycondRp2}). We construct the conditional posterior distribution of $f_{\rm rocky}$ by sampling from  the posterior distribution of $\boldsymbol{\alpha}_2$. We carefully chose the priors on $\boldsymbol{\alpha}_2 \equiv \left\{ R_{\rm mid}, \Delta_R \right\}$ to obtain (nearly) flat priors on the fraction of planets that are rocky at a given $R_p$, $p_0\left( f_{\rm rocky}\left|\right. R_p\right)$; for any $R_p\leq R_{\rm max,rocky}$, the priors on $f_{\rm rocky}$ are flat for $0<f_{\rm rocky}<1$, but there is some ``pile-up" of prior probability at the specific points $f_{\rm rocky}=0$ and $f_{\rm rocky}=1$, corresponding to situations where $R_p$ is larger (smaller) than the maximum (minimum) achieved radius of rocky (non-rocky) planets
\begin{equation}
  p_0\left( f_{\rm rocky}\left|\right. R_p\right) =\frac{1}{4}+ \left(\frac{1}{8} + \frac{R_p}{2R_{\rm max,rocky}}\right)\delta\left(f_{\rm rocky}\right) + \left(\frac{5}{8} - \frac{R_p}{2R_{\rm max,rocky}}\right)\delta\left(f_{\rm rocky}-1\right).
\label{eq:priorfvol2}
\end{equation} 

The conditional posterior distribution $p\left(f_{\rm rocky} \left|\right. R_p, \left\{\boldsymbol{D}_n\right\}_{n=1}^N\right)$ (Figure~\ref{fig:frockycondRp2}) reveals statistically robust insights into the fraction of planets that are sufficiently dense to be rocky, as a function of planet size. 
There is an upper limit on the fraction of large planets ($R_p\gtrsim 1.5~R_{\oplus}$) that are dense enough to be rocky,  stemming from the lack of large dense planets in the sample of {\it Kepler} planets with HIRES RV follow-up. 
There is also a lower limit on the fraction of small planets ($R_p\lesssim 1.5~R_{\oplus}$) that are in the ``potentially rocky" regime, stemming from the detection of a handful of $\sim1.5~R_{\oplus}$ planets that are sufficiently dense to be rocky: Kepler-10b, Kepler-100b (KOI-41.02), Kepler-99b (KOI-305.01), and Kepler-406b (KOI-321.01). The fact that the non-zero lower bound on $f_{\rm rocky}$ extends down to $R_p=0$ is a consequence of the functional form of $f_{2\alpha}$ (Equation~(\ref{eq:fvol2})), which imposes that $f_{2\alpha}\left(R_p\right)$ is monotonically decreasing with planet radius. The posterior distribution of $f_{\rm rocky}$ at a given planet size becomes more uniform (on $f_{\rm rocky}$ between the $R_p$-dependent lower bound and 1) for smaller planet radii as $R_p\to 0$, since the RV-mass constraints on small planets $R_p\lesssim 1~R_{\oplus}$ are weaker and contain less information on the planet compositions than the mass constraints on larger planets. 
The posterior distribution of $f_{\rm rocky}$ at $R_p\sim1.5~R_{\oplus}$ (in the neighborhood of the best-fit radius threshold in the step-function model, $\Delta_R=0$) has significant probability density across the full range of $0\leq f_{\rm rocky}\leq1$.

Our results are largely unchanged when we use a logistic curve instead of a linear function to model a smoother transition between the size regime where most planets are ``potentially rocky" and the size regime where most are volatile-rich  
\begin{equation}
f_{3\alpha} = 1 - \frac{1}{1+ e^{\frac{-2\left(R_p-R_{\rm mid}\right)}{\Delta_{R}}}}.
\label{eq:fvol3}
\end{equation}
\noindent We adopted flat priors on $-0.5R_{\rm max, rock}<R_{\rm mid}<1.5R_{\rm max, rock}$ and $0<\Delta_R<R_{\rm max, rock}$. The constraints on the midpoint of the rocky/non-rocky transition obtained with this logistic curve model are very similar to those obtained for the linear transition: $R_{\rm mid}=1.27^{+0.24}_{-1.24}~R_{\oplus}$, with a mode of $1.48~R_{\oplus}$ and a 95\% upper bound of $1.60~R_{\oplus}$.

Is there evidence in the data for a gradual transition in radius between planets that are sufficiently dense to be rocky, and those that are not? To assess this, we evaluate the Bayesian evidence for each model, 
\begin{equation}
E \equiv \int p\left(\left\{\boldsymbol{D}_n\right\}_{n=1}^N\left|\right.\boldsymbol{\alpha}\right)p\left(\boldsymbol{\alpha}\right) d\boldsymbol{\alpha}
\end{equation}
\noindent The simpler one-parameter step function rocky/non-rocky transition model ($E_1 = 4.1\times10^{-68}$) is mildly favored over the gradual linear transition and logistic transition models ($E_2 = 8.2\times10^{-69}$ and $E_3 = 6.2\times10^{-69}$, respectively). The improvement in the fit in the gradual transition models does not justify the addition of another parameter. This does not mean that there is not some radius range in which both rocky and non-rocky planets co-exist, but rather that more mass-radius measurements of small planets $R_p<2~R_{\oplus}$ are needed to conclusively discern any structure in the transition between rocky and volatile-rich planet populations. 

\subsection{Incident Flux-dependent Rocky/Non-rocky Radius Threshold}
\label{sec:F}

We turn now to exploring whether the transition between planets that are dense enough to be rocky depends on the amount of radiative energy the planet is receiving from its star, $F_p$. We adopt a generalized step-function model, where the radius threshold depends linearly on $\log F_p$,
\begin{equation*}
R_{\rm thresh} \left(F_p\right)= \min\left(R_{\rm thresh0}+R'_{\rm thresh}\log\left(\frac{F_p}{100F_{\oplus}}\right), R_{\rm max,rocky}\right)
\end{equation*}
\begin{equation}
f_{4\alpha}\left(R_p, F_p, R_{\rm thresh0}, R'_{\rm thresh}\right)  \equiv \begin{cases}
       1 &  R_p < R_{\rm thresh}\left(F_p\right)\\
       0 &  R_p\geq R_{\rm thresh}\left(F_p\right).\\
       \end{cases}
           \label{eq:fvol4}
\end{equation}
\noindent In this parameterization, $R_{\rm thresh0}$ is the radius threshold at $F_{p0}=100F_{\oplus}$ (a characteristic median flux for the planets in our sample). We take flat priors on $0<R_{\rm thresh0}<R_{\rm max,rocky}$ and $-R_{\oplus}<R'_{\rm thresh}<R_{\oplus}$.

\begin{figure}
\centering
\includegraphics{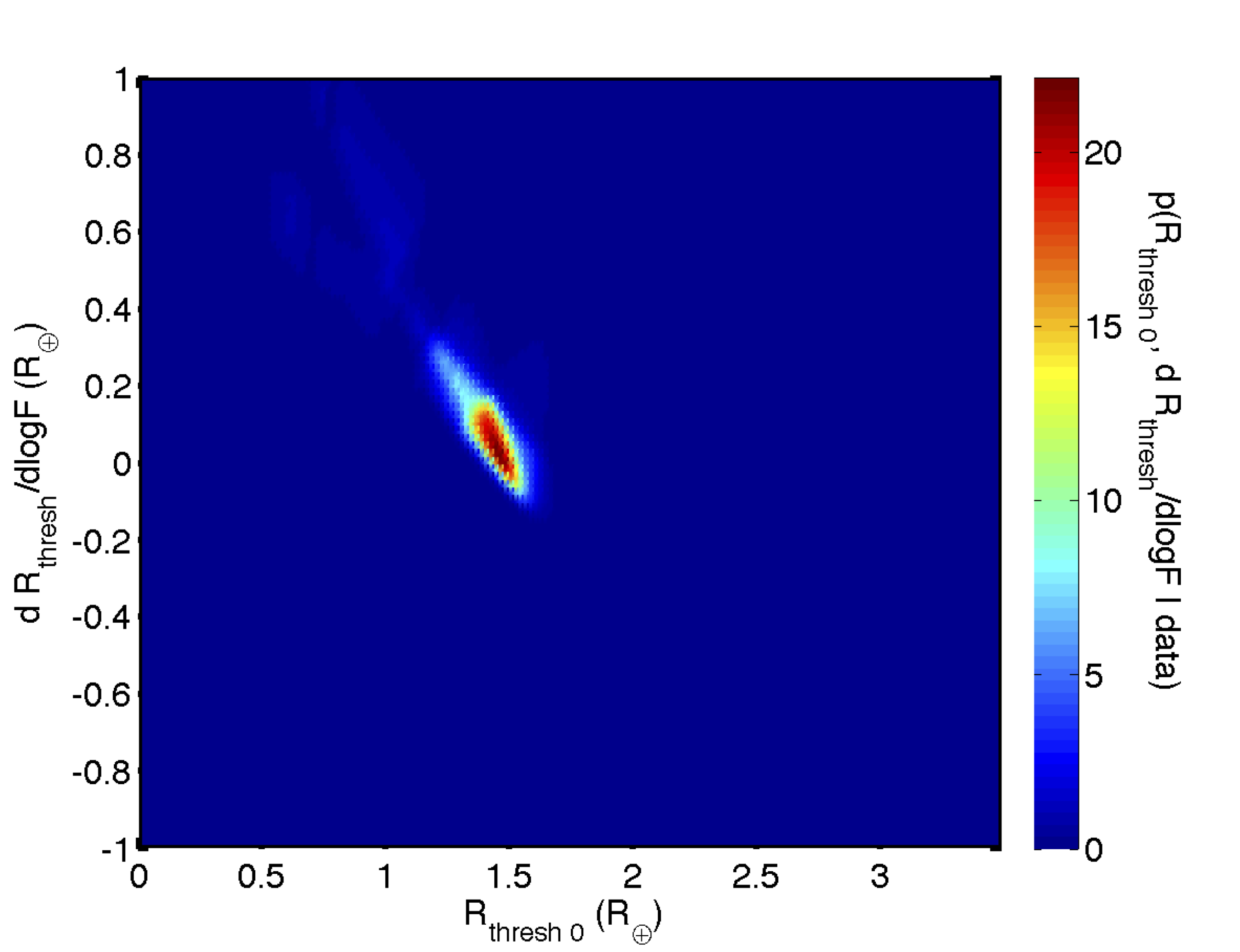}
\caption{ Joint posterior probability density of $R_{\rm thresh0}$ and $R'_{\rm thresh0}$, the parameters in the incident flux-dependent step-function model (Equation~(\ref{eq:fvol4})).}
\label{fig:fvol4} 
\end{figure}

Given the current sample of planets with measured masses and radii, there is no statistically robust evidence for an incident-flux dependence in the radius threshold between planets that are dense enough to be rocky and those that are not. We present the joint posterior PDF of $R_{\rm thresh0}$ and $R'_{\rm thresh0}$ in Figure~\ref{fig:fvol4}. 
The slope of the rocky/non-rocky threshold is consistent with zero (i.e., no dependence on incident flux) and exhibits a slight preference for positive values. Marginalizing over $R_{\rm thresh0}$, we find $R'_{\rm thresh} = 0.11^{+0.35}_{-0.12} R_{\oplus}$. The preference for $R'_{\rm thresh0}>0$ may be expected if more massive planet cores manage to retain their volatiles at higher levels of irradiation.

The Bayesian evidence, $E_4 = 8.3\times10^{-69}$, implies that the flux-dependent rocky/non-rocky transition is less favored than the simple one-parameter step-function model in radius (which does not have any dependence on incident flux). The lack of evidence for an incident flux-dependence in the fraction of planets of a given size that are dense enough to be rocky does not mean that incident flux has no effect on planet compositions.  For sub-Neptune-size planets with H/He envelopes, the planet radius is very sensitive to the planet gas mass fraction, but less sensitive to the total planet mass \citep{RogersEt2011ApJ, Lopez&Fortney2013bApJ}.  Highly irradiated planets may lose their volatile envelopes to atmospheric escape over time, converting larger non-rocky planets (having large $R_p$ with $f_{\rm rocky}\left(R_p\right)\sim0$) into smaller rocky planets (having small $R_p$ with $f_{\rm rocky}\left(R_p\right)\sim1$). To leading order, mass loss would have a more pronounced effect on the radius distribution of planets \citep[the relative number of planets at each radii,][]{Owen&Wu2013ApJ} than on the fraction of planets at specified size that are sufficiently dense to be rocky. 

\section{Discussion}
\label{sec:dis}

\subsection{Sensitivity to the Chosen Planet Sample}
\label{sec:resample}

We have used the {\it Kepler}-discovered planets with Keck-HIRES RV follow-up as our sample to constrain the fraction of planets that are sufficiently dense to be rocky, as a function of planet size. The sample employed in the analysis comprises 27 planet systems. How sensitive are the results to the chosen sample, for instance, to adding or removing a planetary system? Is a single planet system dominating the constraints? 

To assess how sensitive the posterior is to the chosen sample of planet systems, a bootstrapping analysis is used. In each of 1000 iterations, $N=27$ planet systems are sampled with replacement from the original sample of 27 planet-hosting {\it Kepler}+HIRES targets and the hierarchical analysis with the one-parameter step-function model is repeated. Figure~\ref{fig:fvol1} shows the extent to which the posterior probability density of $R_{\rm thresh}$ varies in the bootstrapping analysis; at each value of $R_{\rm thresh}$, the blue shaded region in Figure~\ref{fig:fvol1} encompasses 68.3\% of the bootstrapped values of $p\left(R_{\rm thresh}\left|\right. \rm{data}\right)$. Among the bootstrapped samples, median values of  $R_{\rm thresh}$ span 1.43 to $1.52~R_{\oplus}$ while the 95\% percentile upper bounds on $R_{\rm thresh}$ span 1.54 to $1.67~R_{\oplus}$, where the 34.1\% percentiles above and below the median are quoted. The span of the bootstrapped posterior density of $R_{\rm thresh}$ along with the span of the  median and  95\% percentiles of $R_{\rm thresh}$ do not have a formal meaning in Bayesian statistics (e.g., they should not be interpreted as a credible intervals), but are instead presented to illustrate, in a rough sense, that our results are largely insensitive to adding/removing planet systems from the sample considered.

For many planets in the sample (especially those at $R_p\lesssim1~R_\oplus$), the RV semi-amplitude constraints contain very little information about the planet composition, spanning the range of physically reasonable masses. 
Our results are not sensitive to removing these planets. Eliminating from the analysis of the one-parameter step-function model planets that span $M_p=0$ and $M_{\rm rock,max}\left(R_p\right)$ within their 1~$\sigma$ error bars (namely, KOI 70.04, 70.05, 82.03, 82.04, 82.05, 116.04, 245.02, 245.03, 321.02, 1612.01), a median of $R_{\rm thresh}=1.48~R_{\oplus}$ and 95\% percentile of $R_{\rm thresh}=1.59~R_{\oplus}$ are obtained (identical to the values obtained for the full planet sample to within the quoted precision). We emphasize that to avoid a Malmquist bias toward higher-densities, planets with RV upper limits should be included in the analysis, as we have done. Several RV non-detections in the sample contain valuable information about the planet composition, constraining the planet to be volatile-rich.

\subsection{Threshold Mass--Radius Relations for Rocky Planets}
\label{sec:EarthComp}

We now turn to quantifying the effect of considering a more restrictive range of possible compositions (and hence masses) for rocky planets. 
 The assumptions we have made to date concerning the range of plausible planet masses for rocky planets of a given size are very inclusive.  The pure silicate, and pure iron compositions adopted as the low and high density extremes for rocky planets are extreme end-member scenarios. The photospheres of planet-hosting stars have Fe/Si abundance ratios near 1 (ranging from 0.5 to 1.3) \citep{GrassetEt2009ApJ}. Further, metallic iron and silicates have similar condensation temperatures in the protoplanetary disk \citep[e.g.,][]{Petaev&Wood2005ASPC}, and are expected to concomitantly condense to form solids and to contribute together to the bulk material forming a rocky planet \citep{ValenciaEt2007bApJ}.

Increasing the density (iron fraction) of the limiting low-density composition assumed for rocky planets tends to decrease the inferred fraction of planets of a given size that are sufficiently dense to be rocky. 
This leads to even stronger upper bounds on the planet radius above which most planets are not rocky. 
When we take an Earth-like composition (with 32\% Fe core and 68\% silicate mantle, by mass) as the limiting low-density composition for rocky planets in the one-parameter step-function model we find a median value of  $R_{\rm thresh} = 1.43^{+0.05}_{-0.80}~R_{\oplus}$, and a 95\% confidence upper bound of $R_{\rm thresh} = 1.53^{+0.08}_{-0.04}~R_{\oplus}$ (compared to  $R_{\rm thresh} = 1.48^{+0.04}_{-0.05}~R_{\oplus}$ and a 95\% confidence upper bound of $R_{\rm thresh} = 1.59^{+0.18}_{-0.05}~R_{\oplus}$ assuming a pure-silicate limiting composition). The uncertainties quoted on the percentiles of $R_{\rm thresh}$ span 34.1\% of the values above and below the median evaluated from 1000 bootstrapping samples.
 The threshold mass corresponding to the limiting threshold radius increases to $4.0~M_{\oplus}$ (median) and $5.0~M_{\oplus}$ (95\% upper bound) for an Earth-like composition.

Decreasing the density (iron fraction) of the limiting high-density composition assumed for rocky planets tends (i) to increase the threshold radius of the rocky/non-rocky transition, and (ii) to broaden the constraints on the fraction of planets of a given size that are dense enough to be rocky. Decreasing $M_{\rm rock,max}\left(R_p\right)$ extends the region of planet mass-radius space that is ruled physically implausible in the prior, $p\left(\boldsymbol{\beta}_n\left|\right.\boldsymbol{\alpha}\right)$. Since small-radius solutions are preferentially disfavored, this has the effect of systematically increasing the value inferred for the ``true" radius of each planet relative to that inferred based on the non-informative interior prior, $p_0\left(\beta_n\right)$. The value of $p_{\rm rocky}$ for each planet is also systematically decreased. \citet{MarcusEt2010ApJ} used numerical simulations of giant impacts to compute a minimum radius for iron-rich rocky planets formed by collisional mantle stripping of differentiated planets with an initial 0.33 Fe core mass fraction. Adopting the \citet{MarcusEt2010ApJ} minimum radius relation for rocky planets in the one-parameter step-function model,  we find a median value of  $R_{\rm thresh} = 1.53^{+0.34}_{-0.05}~R_{\oplus}$, and a 95\% confidence upper bound of $R_{\rm thresh} = 1.96^{+0.14}_{-0.36}~R_{\oplus}$.

When the reduced upper and increased lower limits on the density of rocky planets are adopted simultaneously, the location of the rocky/non-rocky threshold shows good agreement with that obtained from the nominal limits (which assume pure iron and pure silicate). Taking the \citet{MarcusEt2010ApJ} minimum radius relation and an Earth-like composition as the limiting maximum radius relation for rocky planets in the one-parameter step-function model,  we find a median value of  $R_{\rm thresh} = 1.48^{+0.07}_{-0.56}~R_{\oplus}$, and a 95\% confidence upper bound of $R_{\rm thresh} = 1.62^{+0.67}_{-0.08}~R_{\oplus}$.

\subsection{Prior Assumptions}

As in any work relying on Bayesian inference, some priors must be chosen. In this section, we discuss our priors and the sensitivity of our results to these choices.

Within the each of the ``potentially rocky" and ``non-rocky" regimes, we have taken a flat prior PDF on the planet mass (Equation~(\ref{eq:flatm})). Throughout the analysis, only the relative weights given to the two regimes at a given $R_p$ has been adjusted; the assumption of a flat prior on the planet mass within each regime has not been varied. 
The most natural alternative to the semi-flat mass prior assumption would be to treat the planet mass as a scale parameter and to adopt a flat prior on $\log{M_p}$. With the RV mass constraints in hand, however, we cannot accurately assess the effect of adopting a flat prior on $\log{M_p}$ in a quantitative way; our importance resampling approach breaks down for the $p\left(M_{nj}\left|\right.R_p\right)\propto1/M_{nj}$ target prior due to insufficient support at low $M_p$ from the interim prior PDF adopted by \citet{MarcyEt2014ApJS}. Qualitatively, however, we expect that adopting a flat prior in $\log{M_p}$ at a given radius would serve to further strengthen our main conclusion that most 1.6~$R_{\oplus}$ planets are not rocky, by adding additional statistical weight at low planet masses (and hence low planet densities). 

In this work, we have parameterized the mass-radius distribution of planets in terms of the fraction of planets or a given size that are rocky, $f_{\rm rocky}\left|\right.R_{p}$. We have assumed simple functional forms for $f_{\rm rocky}\left|\right.R_{p}$ that depend on a few free parameters $\boldsymbol{\alpha}$ as well as on properties of the planet-star system. Exploring four different models for $f_{\rm rocky}\left|\right.R_{p}$, we have shown that our main results are robust against the particular choice of parameterization. The use of low-parameter functional forms for $f_{\rm rocky}$, however, imposes monotonically decreasing and smooth --- except in the step function model, which is discontinuous at $R_{\rm thresh}$ by construction --- behavior on the variations of $f_{\rm rocky}$ with $R_p$. In future work, a more generalized model for the mass-radius distribution of rocky and non-rocky planets could be employed to include more freedom, for example using a step function with $M>1$ steps and a smoothness prior \citep[][Equations (10) and (11)]{HoggEt2010ApJ}, or a Gaussian process. 

Finally, we have assumed flat priors on the population-level parameters $\boldsymbol{\alpha}$ for each model explored. In each case, the posterior PDF of $\boldsymbol{\alpha}$ is significantly narrower than prior probability density, demonstrating that the planet mass-radius data provide stronger constraints on the population-level parameters than do our priors.

\subsection{Insights into Planet Formation}

We have shown that, at planet radii of $1.6~R_{\oplus}$ and above and orbital periods below $\sim50$~days, rocky planets without a gas envelope are less common than planets of the same size with volatile envelopes, at 95\% confidence. 
This largely empirical result is in agreement with the expectations from both core nucleated accretion and rocky planet formation theories. For an Earth-like composition, $1.6~R_{\oplus}$ corresponds to $6~M_{\oplus}$. 

Core nucleated accretion models predict that rocky cores of $6~M_{\oplus}$ imbedded in a gas disk will accrete gas. 
Calculations by \citet{Ikoma&Hori2012ApJ} estimate that a $6~M_{\oplus}$ core could accrete $\gtrsim 0.3\%$ H/He by mass when imbedded in a minimum-mass-solar-nebula protoplanetary disk with a nebular temperature $T_d=550~\mathrm{K}$ and a disk lifetime of 100~kyr. The mass fraction of H/He accreted would be even higher for lower disk temperatures, longer disk dissipation timescales, and less grain opacity. 
Thus, $1.6~R_{\oplus}$ protoplanet cores that assembled before the dissipation of the protoplanetary disk are expected to have acquired a H/He envelope which (if retained) would substantially increase their observed transit radii to $\gtrsim 1.9~R_{\oplus}$. More recent simulations by \citet{Bodenheimer&Lissauer2014ApJ} have revealed that even planets with core masses of $\sim2.2~M_{\oplus}$ may accrete, within 2.0~Myr, $0.037~M_{\oplus}$ of H/He at 0.5~AU and $0.16~M_{\oplus}$ of H/He at 2.0~AU.

Our results constrain the fraction of close-in planets of a specified size that assembled after the dissipation of the protoplanetary gas disk. 
In the standard rocky planet formation scenario \citep[wherein planet embryos grow through runaway and then oligarchic accretion of planetesimals, before finally accumulating into terrestrial planets through a chaotic late-stage of embryo--embryo collisions, see e.g.,][]{RaymondEt2013AstroPh}, the assembly of rocky planets continues after the gas disk has dissipated. Planets formed through this pathway have masses and radii in the high-density ``potentially rocky" regime ($M_{\rm rock,min}\left(R_p\right)\leq M_p\leq M_{\rm rock,max}\left(R_p\right)$) since they would not accrete primary H/He envelopes, and could only accrete water-rich material that is scattered into the terrestrial planet formation region from beyond the snow-line (e.g., by the migration of giant planets). Simulations by \citet{CarterBondEt2012ApJ} predict that most rocky planets retain less than 10 Earth oceans of water delivered by migration (including both surface water and water incorporated into the mantle). 
Formation of massive rocky planets $\left(6~M_{\oplus}\right)$ would require a disk with a high surface mass density of solids (relative to the minimum mass solar nebula). The characteristic mass-scale of rocky planets (derived by assuming that each planet accretes all of the condensed material in an annulus centered on its orbit of width proportional to the planet Hill sphere) scales with the disk surface mass density in solids, $\sigma$, as $M_p\propto a^3\sigma^{3/2}M_{\star}^{-1/2}$ \citep{Lissauer1995Icarus}. Further, $N$-body simulations of the giant impact stage of rocky planet formation have found that the mass of the most massive planet formed scales nearly linearly with the total mass in protoplanets \citep{KokuboEt2006ApJ}. 
While our result that most close-in planets larger than $1.6~R_{\oplus}$  are not rocky is not necessarily surprising, this is the first time that sufficient mass-radius constraints for sub-Neptune-size planets exist to extract population level composition constraints in a statistically robust way.

Current RV follow-up of close-in {\it Kepler} planets does not rule out nor definitively rule in the possibility small planets with substantial complements of low-density astrophysical ices. 
Improving the constraints on the density distribution of small planets on close orbits may help to resolve whether the compact close-in systems of low-density planets discovered by {\it Kepler} formed in situ \citep[e.g,][]{Hansen&Murray2012ApJ, Ikoma&Hori2012ApJ, Chiang&Laughlin2013MNRAS}, or alternatively acquired their volatiles farther out in the disk and migrated in to their current locations \citep[e.g,][]{RogersEt2011ApJ, SwiftEt2013ApJ, CossouEt2014A&A}.
Low density condensables (water and other astrophysical ices) are a tracer a planet's formation location: planets formed beyond the snow line are expected to initially contain an ice mass fraction comparable to its rock mass fraction \citep{Lewis1972Icarus}, while planets formed on the close-in orbits ($P_{orb}<50~{\rm days}$, representative of the {\it Kepler} planets with RV follow-up) are expected to only have trace amounts of condensables \citep[e.g.,][Table 1]{RaymondEt2008MNRAS}. 
Small radius planets on close-in orbits would loose any envelopes of light gases (H and He) on short timescales ($\lesssim 1~\rm{Myr}$) \citep{RogersEt2011ApJ, Lopez&Fortney2013ApJ}. Thus, a small planet with a bulk density below that of silicate ($3.3~{\rm g\, cm^{-3}}$ uncompressed or $5.6~{\rm g\, cm^{-3}}$ mean compressed density at $1.5~R_{\oplus}$) would be a clear signature of astrophysical ices and an initial formation location beyond the snow line.
Stronger RV detections and upper limits on the masses of planets with radii ($R_p<1.5~R_{\oplus}$) are needed. 

Our analysis focusses on the current composition of planets observed today. We expect that the fraction of planets of a given radius that are rocky was even smaller in the past, since volatile loss processes outpace volatile sources. Close-in sub-Neptunes lose volatiles over time to photo-evaporation \citep[e.g.,][]{LeCavelierDesEtangs2007A&A, ValenciaEt2010A&A, RogersEt2011ApJ, BoueEt2012A&A, LopezEt2012ApJ}. In contrast, mechanisms to replenish a lost envelope are not predicted to provide volatiles in quantities sufficient to substantially increase the transit radius. Rates of volcanism per unit mass on rocky Earth-like exoplanets are not expected to exceed 10 times that of the present-day Earth \citep[$1.7\times10^{-11}~\rm{yr^{-1}}$,][]{Best&Christiansen2001book} for planets older than 2~Gyr \citep{KiteEt2009ApJ}. Further, late delivery of volatiles by impacting comets will not contribute sufficient volatiles to produce an observable effect on the transit radius, while large impactors \citep[with diameters larger than the atmospheric scale height, e.g.,][]{Ahrens1993AREPS} will erode the planet atmosphere. 

\subsection{The Nature of sub-Neptune-size {\it Kepler} Planet Candidates}

Our hierarchical Bayesian analysis gives insights into the nature of the  thousands of transiting {\it Kepler} planet candidates that do not have measured masses. 
Based on the sample of {\it Kepler} planets with RV follow-up, we found that most planets larger than $1.6~R_{\oplus}$ are so low-density that a volatile envelope must contribute significantly to their transit radius. The {\it Kepler} Mission developed a working nomenclature for planets, based solely on their radii; describing planets $<1.25~R_{\oplus}$ as Earth-size, 1.25--$2.0~R_{\oplus}$ as Super Earth-size, and  2--$6~R_{\oplus}$ as Neptune-size \citep[e.g.,][]{BoruckiEt2011ApJ}. Our results (Figure~\ref{fig:frockycondRp2}) provide quantitative estimates of the fraction of planets in each of these ranges that are sufficiently dense to be rocky. One of the primary science goals of the {\it Kepler} mission is to calculate the occurrence rate of Earth-like planets in the habitable zones of sun-like stars, $\eta_{\oplus}$. We suggest that the operational definition of ``Earth-like" focus on planets with $R_p\lesssim 1.6~R_{\oplus}$, to consider planets with a significant probability of having a rocky composition. 

The limits on the fraction of planets of a given size that are dense enough to be rocky derived in this work should be regarded as upper bounds; it is likely that a smaller fraction of planets of any size are rocky.  We have specifically investigated the fraction of planets that are sufficiently dense to be rocky (i.e. more dense than an iron-poor, pure silicate composition). Planets sufficiently dense to be rocky may still harbor a thick envelope of volatiles that contributes to its transit radius, if the volatiles are offset by a more iron-rich make-up for the rocky-component of the planet.

Our analysis does not preclude the possibility of large rocky planets. We have assumed a smooth functional for how the fraction of planets that are dense enough to be rocky depends on planet size. With the current sample of planet mass and radius measurements, we do not capture complex structures in the planet mass-radius distribution. Massive rocky planets larger than 1.6~$R_{\oplus}$ on close orbits may still exist, but they are the exception rather than the rule.

The current sample of transiting {\it Kepler} planets with RV follow-up is limited to planets on close-in orbital periods ($P<50~\rm{days}$). 
We might naively anticipate that the fraction of planets that are rocky will decrease at greater separations from the host star. 
At longer orbital periods, planets are less irradiated by their stars and will experience less photoevaporation. Additionally, the temperature in the protoplanetary disk decreases with distance from the star, making it easier for planetary embryos farther out to accrete primordial H/He envelopes. Reality is likely more complicated, however, and this naive expectation may break down. For instance, if the close-in compact systems of volatile-rich low-mass/low-density planets discovered by {\it Kepler} formed by Type I inward migration that stalled at the inner edge of the gas disk, rocky planets formed after the dissipation of the gas disk could be a more important fraction of the planet population further out $\left(\sim1~{\rm AU}\right)$.
Exoplanetary science is a data-driven field, and we ultimately must push to measure the masses and radii of planets at longer orbital periods to get a clearer picture of small planet composition demographics and to extrapolate $f_{\rm rocky}$ to the habitable zone.

Kepler-22b, the first habitable zone planet with measured radius \citep[ $2.4~R_{\oplus}$][]{BoruckiEt2012ApJ}, is an illustrative example of how the sample of {\it Kepler} planets with HIRES RV follow-up affects our prior assumptions on the nature of transiting planets with unknown masses. When it was first announced in 2012, the RV mass upper limit on Kepler-22b (36, 82 and $124~M_{\oplus}$ $1\sigma$, $2\sigma$ , and $3\sigma$ upper limits) did not substantially constrain the composition, and a rocky planet scenario for Kepler-22b was considered a serious possibility \citep{BoruckiEt2012ApJ}. Even with subsequent RV follow-up that has further constrained Kepler-22b's mass (42, 54 and $66~M_{\oplus}$ $1\sigma$, $2\sigma$ , and $3\sigma$ upper limits) the nature of Kepler-22b is ambiguous; based on its individual mass-radius constraints $p_{\rm rocky}=0.75$. Placed in the context of the other $\sim2.4~R_{\oplus}$ planets with mass measurements (all of which are constrained to have low density), we find that Kepler-22b is likely not rocky, instead having a volatile envelope that contributes significantly to its volume. Within the constraints of the linear-transition model, $f_{\rm rocky}\left(2.4~R_{\oplus}\right)<0.02$ at higher than 95\% confidence.

Kepler-10c is another notable planet in our sample of {\it Kepler}-discovered planets with HIRES RV mass constraints.  
Intensive HARPS-N RV-follow-up measured the planet mass to be $M_p=17.2\pm1.9~M_{\oplus}$ \citep{DumusqueEt2014ApJ}. 
Even with its high reported density $\left(\rho_p=7.1\pm1.0~\rm{g\,cm^{-3}}\right)$, Kepler-10c mass and radius are inconsistent with a rocky composition by more than $1~\sigma$; it must have a substantial volatile envelope (of astrophysical ices or H/He) that contributes to its transit radius. Based on the mass and radius quoted from \citet{DumusqueEt2014ApJ}, $p_{\rm rocky}\sim0.1$; Kepler-10c is likely not rocky nor solid. This is consistent with our finding that at a radius of $2.3~R_{\oplus}$, rocky planets are rare relative to low-density volatile-rich planets.

\section{Conclusions}
\label{sec:con}

We have developed a hierarchical Bayesian approach to constrain the fraction of planets that are sufficiently dense to be rocky as a function of planet size from a sample of transiting planets with mass constraints. Applying this approach to the sample of {\it Kepler} planets with Keck-HIRES RV follow-up, we have shown that at any radius equal to or larger than $1.62~R_{\oplus}$, the majority (50\% or more) of planets of that size are too low density to be comprised of Fe and silicates alone (at 95\% statistical confidence). 
With the current sample of {\it Kepler} planets having Keck HIRES RV follow-up, we can neither distinguish between an abrupt transition and gradual transition from rocky to ``volatile-shrouded" planets, nor conclusively identify a dependence on planet irradiation. More planet mass-radius measurements with smaller error bars and well quantified selection effects are needed to constrain the structure of the transition in between rocky and non-rocky planets. 

Our constraints on the radii above which most planets are too low density to be composed of iron and silicate alone provide a useful empirical constraint for planet formation theories. These results give insights into the masses and compositions of the remaining sub-Neptune size {\it Kepler} planet candidates that orbit stars which are too faint for RV follow-up, and motivate an operational definition of ``Earth-like" that can be used to calculating the occurrence rate of Earth-analogs, $\eta_\oplus$. Our conclusions are the result of a largely model-independent statistical interpretation of empirical data. The only planet interior structure models that entered into our analysis were those defining the limiting low-density and high-density mass-radius relations for rocky planets. 

With larger numbers of planets having constraints on both their mass and radius, one could eventually include extra parameters in the model to constrain the mass-radius distribution of planets as a function of planet orbital period/incident flux, stellar mass, and planet multiplicity. Each new extra dimension of characterization promises to yield additional insights into planet formation and evolution. For the current sample of planet mass-radius measurements, a more sophisticated treatment of selection effects in the analysis of the current sample of planet masses and radii may provide an avenue to move beyond exploring conditional mass/composition distributions at specific planet radius, to the joint mass-radius-incident flux distribution. Modeling the selection effects will be messy due to the ever-evolving criteria applied to select {\it Kepler} planet candidates for RV follow-up, however, and will be the subject of a future paper. 

The future is bright for small planet mass-radius measurements. Continued Doppler monitoring of {\it Kepler} planets with Keck-HIRES and HARPS-N will improve the planet mass constraints. GAIA will be of great help by measuring the distances to {\it Kepler} targets and thereby reducing the uncertainties on the host star and planet radii. The Transiting Exoplanet Survey Satellite (TESS), scheduled to launch in 2017, will find transiting planets around bright stars that are more amenable to RV follow-up than the typical {\it Kepler} target.
Moving forward into the TESS-era, adopting an algorithmic approach with pre-determined criteria to select TESS transiting planet candidates for RV follow-up would facilitate subsequent statistical inferences about the underlying composition distribution of planets from the measured masses and radii.  Such a survey strategy would help to leverage as much information as possible from RV marginal detections and RV non-detections. Improving constraints on the composition distribution of small planets would benefit from devoting a pre-determined fraction of TESS RV follow-up time to a mass-radius survey that is allocated separately and therefore buffered against the inevitable competing pressures to follow-up high-impact individual systems. 

The hierarchical Bayesian model approach that we have outlined for constraining the composition distribution planets has several strengths: (i) it directly couples interior structure models to the the mass radius posterior distributions output from the analysis of transit and RV data; (ii) it uses the information contained in marginal detections and non-detections; and (iii) it can naturally be extended to account for survey selection effects. These features of our model approach will become even more crucial as planet hunters continue to push toward smaller, less-massive planets at longer orbital periods near the sensitivity limits of the RV and transit techniques.

\acknowledgments

I would like to especially thank Geoff Marcy for extensive discussions and advice, Howard Isaacson for sharing samples from his MCMC fits to the Keck RV data, and an anonymous reviewer whose comments helped to improve this manuscript.  I am also grateful for helpful discussions with Lauren Weiss, Rebekah Dawson, Tim Morton, Brice-Olivier Demory, Brian Jackson, Jack Lissauer, Tom Loredo, Eric Ford, David Hogg, and John Johnson. 
I acknowledge support provided by NASA through Hubble Fellowship grant \#HF-51313 awarded by the Space Telescope Science Institute, which is operated by the Association of Universities for Research in Astronomy, Inc., for NASA, under contract NAS 5-26555. This work also benefited from the Summer Program on Modern Statistical and Computational Methods for Analysis of {\it Kepler} Data, held at SAMSI, Research Triangle Park, NC in 2013 June. This research has made use of the Exoplanet Orbit Database and the Exoplanet Data Explorer at exoplanets.org.

\bibliography{apj-jour,exoplanets}

\bibliographystyle{apj}

\clearpage

\end{document}